\documentclass[pdftex,twocolumn,epjc3]{svjour3}          

\RequirePackage[T1]{fontenc}

\smartqed  

\RequirePackage{graphicx}
\RequirePackage{mathptmx}      
\RequirePackage{flushend}
\RequirePackage[numbers,sort&compress]{natbib}
\RequirePackage[colorlinks,citecolor=blue,urlcolor=blue,linkcolor=blue]{hyperref}
\usepackage{amsmath}
\journalname{Eur. Phys. J. C}

\newcommand{\avrpT}{\ensuremath{\langle p_t\rangle \mathrm{\; vs\; } N_{ch}}}

\begin{document}

\title{Improving the Simulation of Quark and Gluon Jets with Herwig 7}


\author{Daniel Reichelt\thanksref{e1,addr1} \and Peter Richardson\thanksref{e2,addr2,addr3} \and Andrzej Siodmok\thanksref{e3,addr4}}

\thankstext{e1}{e-mail: daniel.reichelt@tu-dresden.de}
\thankstext{e2}{e-mail: peter.richardson@durham.ac.uk}
\thankstext{e3}{e-mail: andrzej.siodmok@ifj.edu.pl}

\institute{Institut f\"ur Kern- und Teilchenphysik, Technische Universit\"at Dresden\label{addr1}
          \and
          Theory Department, CERN, Geneva\label{addr2}
          \and
          IPPP, Department of Physics, Durham University\label{addr3}
          \and
          The Henryk Niewodnicza\'nski Institute of Nuclear Physics in Cracow, Polish Academy of Sciences\label{addr4}
}

\date{Received: date / Accepted: date}

\maketitle

\begin{abstract}
  The properties of quark and gluon jets, and the differences between them, are
  increasingly important at the LHC. However, Monte Carlo event generators are normally
  tuned to data from $e^+e^-$ collisions which are primarily sensitive to quark-initiated jets.
  In order to improve the description of gluon jets we make improvements to the perturbative and 
  the non-perturbative modelling of gluon jets and include data with gluon-initiated jets
  in the tuning for the first time. The resultant tunes significantly improve the
  description of gluon jets and are now the default in \textsf{Herwig}~7.1.
\end{abstract}

\section{Introduction}
\label{intro}

Monte Carlo generators are essential tools, both for the design
of future experiments and the analysis of data from the LHC, and previous
collider experiments. Modern event generators \cite{Bellm:2015jjp,Sjostrand:2014zea,Gleisberg:2008ta}
provide a simulation of exclusive events\linebreak based on the
combination of fixed-order perturbative results, resummation of
large logarithms of scales using the parton-shower approach and
non-perturbative models of hadronization and multiple-parton scattering.\footnote{For a recent review of modern Monte Carlo event generators see \cite{Buckley:2011ms}.}

These simulations rely on universality and factorization in order to 
construct a simulation of the complex final states observed in hadronic 
collisions. This allows the simulation of final-state radiation in the parton
shower and the non-perturbative hadronization models to be first developed,
and the parameters of the the model tuned, using the simpler and cleaner
environment of $e^+e^-$ collisions, and then applied to more
complicated hadronic collisions. These models are then combined with
the parton-shower simulation of initial-state radiation, a multiple
scattering model of the underlying event and a non-perturbative colour
reconnection model
in order to describe hadronic collisions.
In principle universality requires that the colour reconnection model
is also used to describe leptonic collisions. In practice however 
colour reconnection has little
effect on the distributions which so far have been used to develop 
and tune the models. These models are therefore usually either not
included at all for the simulation of leptonic collisions, or if they
are, the parameters are determined by tuning to hadronic data sensitive
to multiple partonic scattering.

As the LHC accumulates data at an unprecedented rate there are a number of observables
which are not well described by current Monte Carlo event generators, and where
the limitations of
this approach have started to become obvious, 
for example:
\begin{itemize}
\item the difference in the properties of jets initiated by quarks and gluons
is not well described with generators predicting either a larger or smaller difference between
the jets than is observed by the LHC experiments \cite{Aad:2014gea};
\item the transverse momentum spectra of identified baryons and strange hadrons
  which are not well described by current generators. \cite{Khachatryan:2011tm};
\item long-range correlations in high multiplicity events\cite{Khachatryan:2010gv,Aad:2015gqa}.
\end{itemize}
In this paper we will focus on improvements to the perturbative and non-perturbative modelling
to give a better description of both quark- and gluon-initiated jets, as well as the differences
between them in \textsf{Herwig 7}. Beyond leading order
there is no clear distinction between quark and gluon jets and the definition will depend on the
analysis.\footnote{See Ref.\,\cite{Gras:2017jty} for a more detailed discussion.}
As $e^+e^-$ annihilation to hadrons starts 
with an initial partonic quark-antiquark configuration the data used
to develop the final-state parton-shower algorithm, tune its parameters
and those of the hadronization model, are dominated by quark-initiated jets.
However at the LHC jets initiated by gluons can often dominate,
depending on the production process, rapidity and transverse momentum of the jets.
Regrettably while there is great interest in the differences between quark and gluon jets
at the LHC most of the experimental studies have concentrated on differentiating between quark and
gluon jets using neural network, or similar, techniques which makes a direct comparison
with simulated hadron-level events impossible. We will therefore use
some recent data from the ATLAS experiment~\cite{Aad:2016oit} which is sensitive to 
both quark and gluon jet properties, together with data on gluon jets in $e^+e^-$
collisions from the OPAL experiment~\cite{Abbiendi:2003gh,Abbiendi:2004pr} which has
not previously been used in the development and tuning of the current generation
of Monte Carlo event generators to study the properties of gluon jets.

In the next section we will first recap the default parton-shower algorithm used in \textsf{Herwig\ 7}
focusing on recent changes we have made to improve the simulation of both
quark and gluon jets. In Section~\ref{sec:hadron} we will briefly review the
important parameters in the cluster hadronization model used in \textsf{Herwig\ 7}
and identify the issues which may lead to different treatments of quark and gluon jets.
We will then discuss the tuning strategy used to produce the tunes presented in this paper.
We present our results in Section~\ref{sec:results}\footnote{Additional results on 
quark and gluon jet discrimination power are included in the Appendix.} followed by our conclusions.

\section{Herwig 7 Parton-Shower Algorithm}
\label{sec:shower}
The default \textsf{Herwig\ 7} parton-shower algorithm \cite{Gieseke:2003rz} is an improved angular-ordered
parton shower. In this approach the momenta of the partons produced in the parton shower are decomposed 
in terms of the 4-momentum of the parton initiating the jet, $p$ ($p^2=m^2$,
the {\em on-shell} parton mass-squared), a light-like reference vector, 
$n$, in the direction of
the colour partner of the parton initiating the jet 
and the momentum transverse to the direction of $p$ and $n$. The four momentum of 
any parton produced in the evolution of the jet can be decomposed as
\begin{equation}
q_i = \alpha_i p + \beta_i n + q_{\perp i},
\end{equation}
where $\alpha_i$ and $\beta_i$ are coefficients and $q_{\perp i}$ is the transverse four momentum 
of the parton ($q_{\perp i}\cdot p = q_{\perp i}\cdot n =0$).

\begin{figure}
\begin{center}
\includegraphics{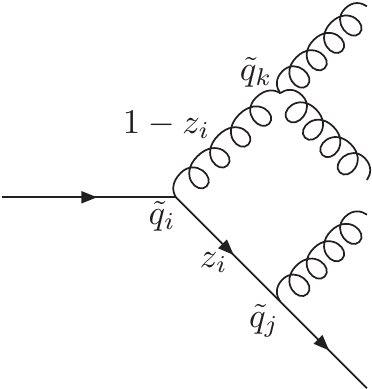}
\end{center}
\caption{Branching of the parton $i$ to produce the partons $j,k$ which then undergo subsequent branching.}
\label{fig:splitting}       
\end{figure}

If we consider the branching of a final-state parton $i$ to two partons $j$ and $k$, i.e. $i\to j k$ as shown in Fig.\,\ref{fig:splitting}, the branching
is described by the evolution variable
\begin{equation}
\tilde{q}^2_i = \frac{q^2_i-m^2_i}{z_i(1-z_i)}, 
\end{equation}
where $q^2_i$ is the square of the virtual mass developed by the parton $i$ in the branching, $m_i$ is the physical mass of parton $i$, and $z_i$
is the momentum fraction of the parton $j$ defined such that 
\begin{equation}
\alpha_j = z_i \alpha_i, \ \ \ \ \ \ \ \ \ \ \ \ \ \ \ \ \ \ \ \alpha_k = (1-z_i)\alpha_i.
\end{equation}
The transverse momenta of the partons produced in the \linebreak branching are
\begin{equation}
q_{\perp j} =    z_i q_{\perp i} + k_{\perp i}\ \ \ \ \ \ \ \ \ 
q_{\perp k} = (1-z_i)q_{\perp i} - k_{\perp i},
\end{equation}
where $k_{\perp i}$ is the transverse momentum generated in the \linebreak branching.
In this case the virtuality of the parton $i$ is
\begin{equation}
q_i^2 = \frac{p^2_{Ti}}{z(1-z)} + \frac{m_j^2}z+\frac{m_k^2}{1-z},
\label{eqn:mass}
\end{equation}
where $p_T$ is the magnitude of the transverse momentum produced in the branching
defined such that \mbox{$k_{\perp i}^2=-p^2_{Ti}$}.

In this case the probability for a single branching to happen is
\begin{equation}
{\rm d} \mathcal{P} = \frac{{\rm d}\tilde{q}^2_i}{\tilde{q}^2_i}\frac{\alpha_S}{2\pi}
\frac{{\rm d} \phi_i}{2\pi} {\rm d} z_i P_{i\to jk}(z,\tilde{q}),
\label{eqn:prob}
\end{equation}
where $P_{i\to jk}(z,\tilde{q})$ is the quasi-collinear splitting function,
and $\phi_i$ is the azimuthal angle of the transverse momentum $k_{\perp i}$
generated in the splitting.

As the branching probability is singular for massless partons an infrared cut-off
is required to regularise the singularity. In
\textsf{HERWIG\ 6}~\cite{Corcella:2000bw} and early versions
of \textsf{Herwig++}~\cite{Gieseke:2003hm} the cut-off was implemented by giving the partons an
infrared mass. However while this remains an option in later versions of
\textsf{Herwig++} and \textsf{Herwig 7}~\cite{Bellm:2015jjp} the default cut-off is now on the minimum
transverse momentum of the branching~\cite{Bahr:2008pv}.

In order to resum the dominant subleading\linebreak logarithms~\cite{Catani:1990rr}
the transverse momentum of the branching is used as the scale for
the strong coupling constant. This also means that the strong
coupling used in the parton shower is that defined in the Catani-Marchesini-{\linebreak}Webber~(CMW)
scheme which includes the subleading\linebreak terms via a redefinition of QCD scale,
$\Lambda_{\rm QCD}$.

While this specifies both the branching probability and kinematics of the partons for a single emission 
in the case of subsequent emission from the daughter partons $j$ and/or $k$ we must decide
which properties of the originally generated kinematics to preserve once the
masses of $j$ and/or $k$ in Eqn.\,\ref{eqn:mass} are no longer the infrared cut-off masses
but the virtualities generated by any subsequent emissions.
While this choice is formally subleading it can have a large effect on physical observables.

In \textsf{Herwig++} the transverse momentum of 
the branching was calculated using Eqn.\,\ref{eqn:mass} and the infrared cut-off\linebreak masses when 
the emission was generated and then preserved during the subsequent evolution of the daughter partons.
In \textsf{Herwig 7.0} the default option was to instead preserve the virtuality of the branching
and calculate the transverse momentum
of the branching using the virtual masses the daughter partons develop due to 
subsequent emissions. This means that if the daughter partons
develop large virtual masses the transverse momentum of the branching is reduced,
and in some cases the branching has to be vetoed if there is no solution of
Eqn.\,\ref{eqn:mass}. However, 
this choice inhibits further soft emission and significantly changes
the evolution by vetoing emissions and leads to incorrect evolution of observables.
We therefore consider a further choice in which 
if it is possible to preserve the virtuality and still have a solution for $p_T^2>0$
we do so, however if this is not possible instead of vetoing the emission we set
$p_T=0$ and allow the virtuality to increase.

The most important parameters which affect the\linebreak behaviour of the parton shower
and which we will tune in this paper are:
\begin{itemize}
\item the choice of whether to preserve $p_T$ or $q^2$ during the subsequent evolution;
\item the value of the strong coupling constant \textbf{AlphaMZ},\linebreak taken to be
$\alpha^{\rm CMW}_S(M_Z)$, value of the coupling
  constant in the CMW scheme at the mass of the $Z$ boson, $M_Z$;
\item the cut-off in the parton shower\footnote{There is an option to extend 
the parton-shower radiation to the non-perturbative region and effectively remove the cut-off, 
see~\cite{Gieseke:2007ad}.}.
  For a cut-off in $p_T$ this is the minimum transverse momentum allowed for the branchings in the shower, $p_T^{\min}$.
  For a virtuality cut-off we parameterize the threshold for
      different flavours as
\begin{equation}
  Q_{g}=\max\left(\frac{\delta-am_{q}}{b},c\right),
\end{equation} 
where $a$ and $b$ are parameters
chosen to give a threshold which is slightly reduced for heavier quarks. The parameter
$c=0.3$ GeV is chosen to prevent the cutoff becoming too small, we also keep 
the default value of $b=2.3$.
Only the parameters $\delta$~(cutoffKinScale) and $a$~(aParameter) are tuned to the data.
\end{itemize}

\begin{figure}
\begin{center}
\includegraphics[width=0.5\textwidth]{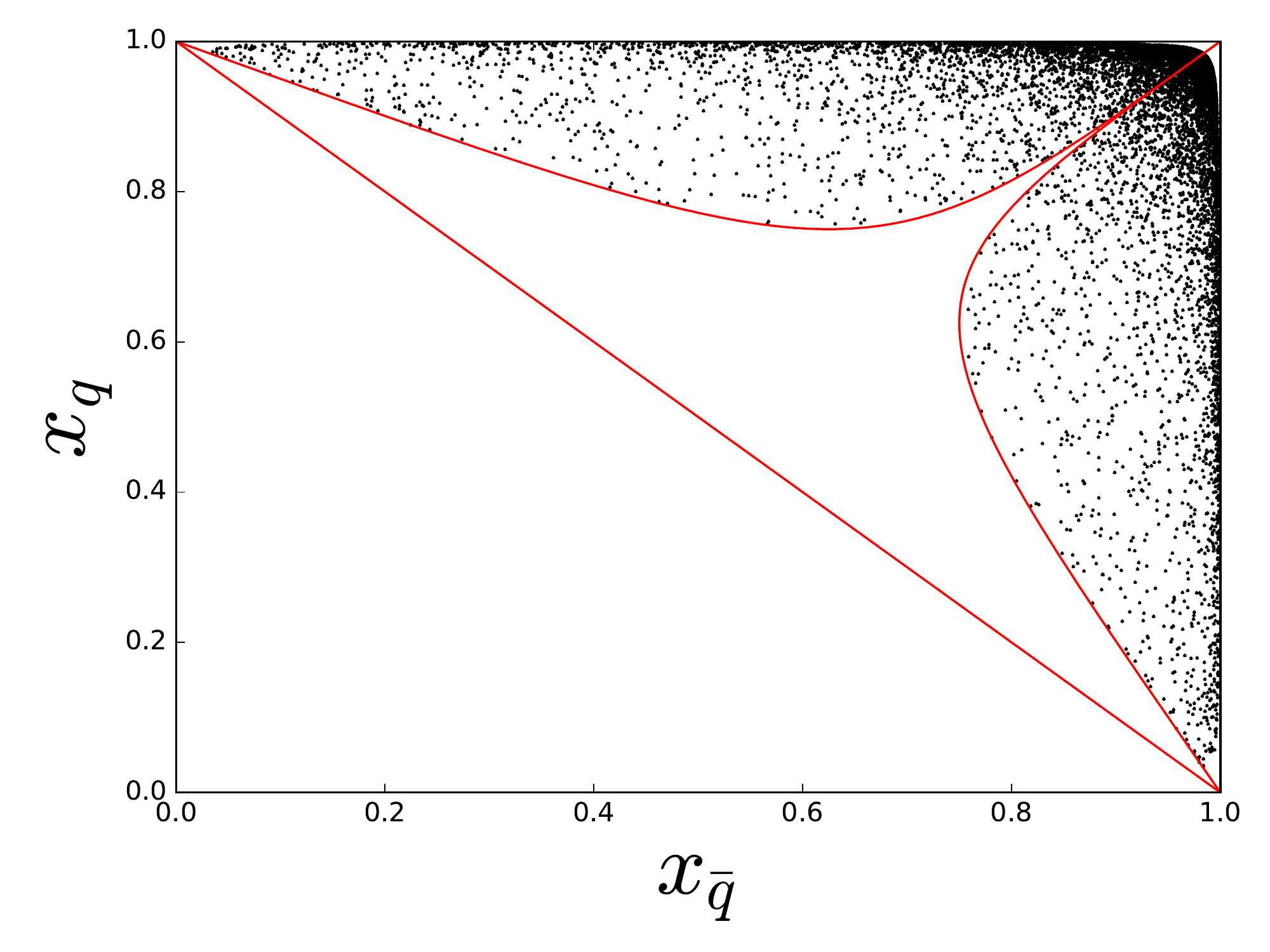}
\end{center}
\caption{Dalitz plot for $e^+e^-\to q \bar q g$ showing the region of phase space filled
by one emission from the quark and antiquark in the angular-ordered parton shower. The line shows
the limits for the parton-shower emission. $x_i = 2 E_i/Q$ where $E_i$
is the energy of parton $i$ and $Q$ is the centre-of-mass energy of the collision.}
\label{fig:dalitzShower}
\end{figure}

\begin{figure}
\begin{center}
\includegraphics[width=0.5\textwidth]{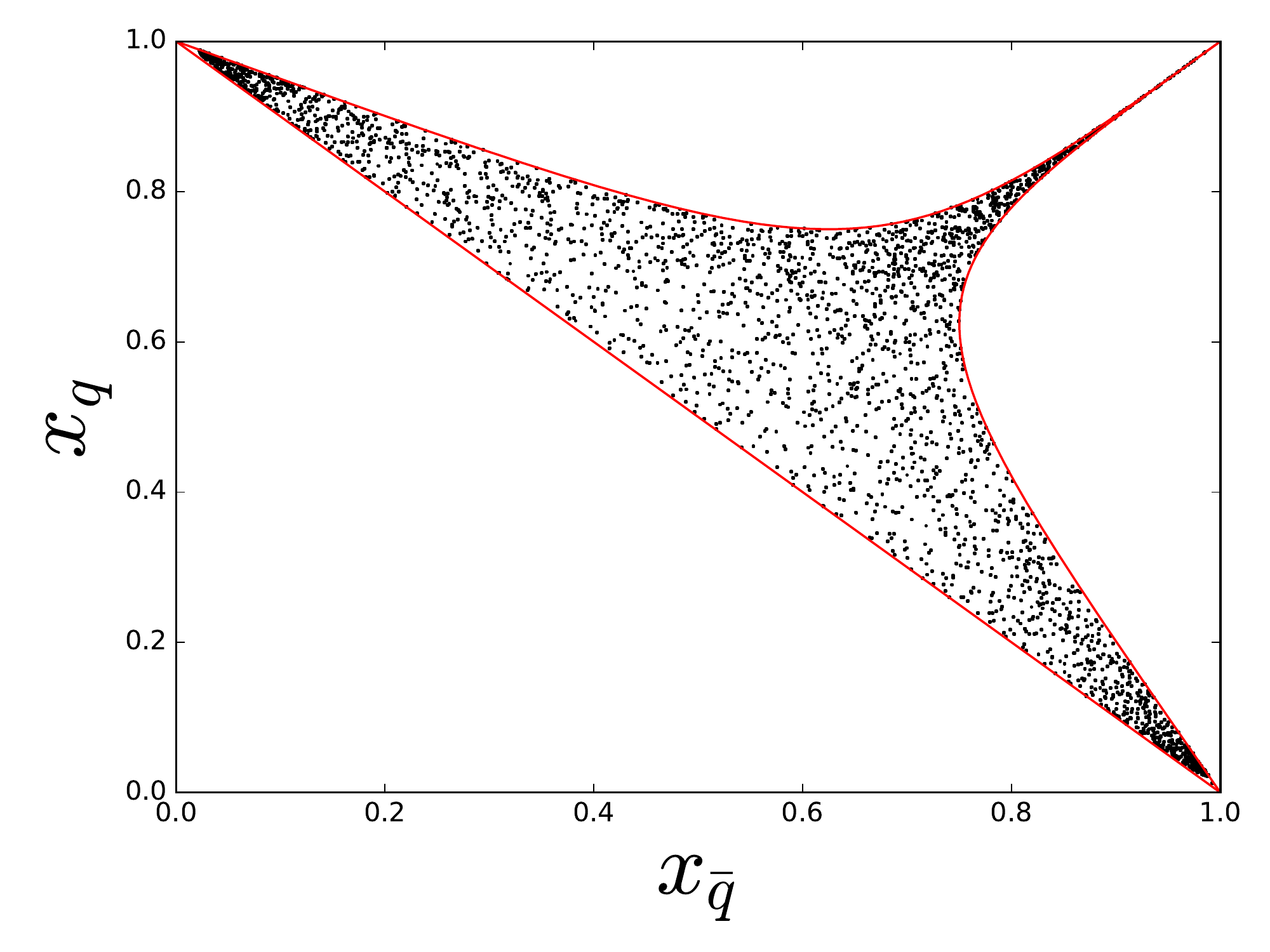}
\end{center}
\caption{Dalitz plot for $e^+e^-\to q \bar q g$ showing the emission from the hard matrix
element correction into the {\it dead-zone} which is not populated by parton-shower emission.
The line shows the limits for the parton-shower emission. $x_i = 2 E_i/Q$ where $E_i$
is the energy of parton $i$ and $Q$ is the centre-of-mass energy of the collision.}
\label{fig:dalitzHard}
\end{figure}

There is one other major feature of the angular-ordered parton shower which we
need to consider. The angular ordering of the parton shower, which is used to 
implement the phenomenon of colour coherence, leads to regions of phase space in
which there is no gluon emission. Consider for example the process
$e^+e^-\to q \bar{q} g$. In this case there is a {\it dead-zone} which is
not filled by one emission from the parton shower, as shown in Fig.\,\ref{fig:dalitzShower}.
Given this deficit of hard, wide-angle emission it is necessary to combine
the parton-shower with the fixed-order calculation of $e^+e^-\to q\bar q g$. There are now
a range of techniques which can achieve this including both the next-to-leading order
normalization of the total cross section, or including the fixed-order results for
multiple emissions. However, for
our purposes it is sufficient to consider the simplest matrix-element correction
approach where the {\it dead-zone} is filled using the leading-order matrix element
for  $e^+e^-\to q\bar q g$, as shown in Fig.\,\ref{fig:dalitzHard}, together with
the reweighting of emission probability, Eqn.\,\ref{eqn:prob}, to the exact 
leading-order result, for any emission which could have the highest transverse momentum in
the parton shower.\footnote{Due to the choice of ordering variable the hardest emission may not be 
the one that has the highest value of the ordering variable, i.e. the hardest emission may be 
not the first emission.}

\begin{figure}
\begin{center}
\includegraphics[width=0.5\textwidth]{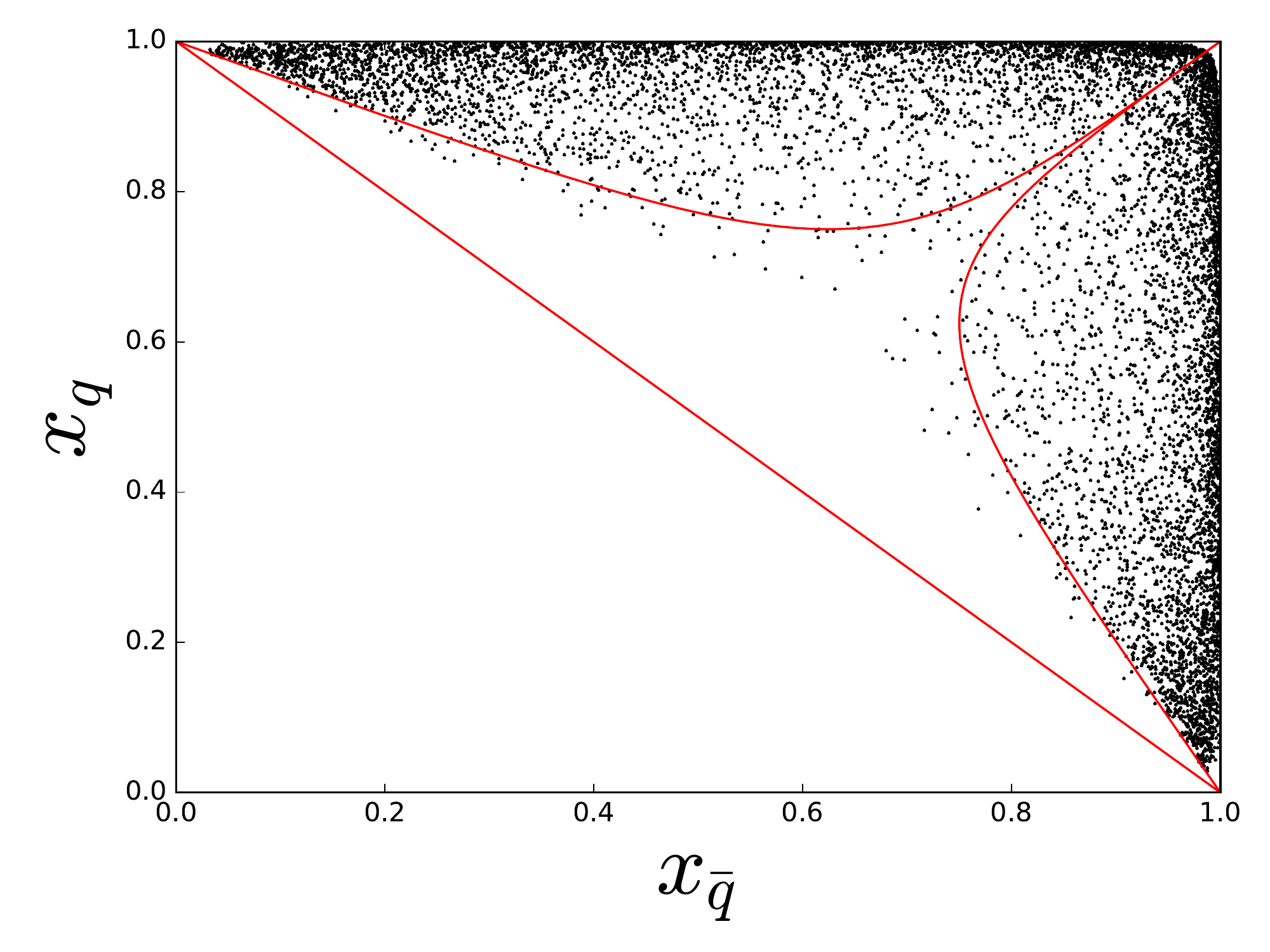}
\end{center}
\caption{Dalitz plot for $e^+e^-\to q \bar q$ showing the region of phase space filled
after multiple emission from the quark and antiquark in the angular-ordered parton shower.
The transverse momentum of the branchings was preserved in the case of multiple emission.
The line shows the limits for the parton-shower emission for a single emission. $x_i = 2 E_i/Q$ where $E_i$
is the energy of parton $i$ and $Q$ is the centre-of-mass energy of the collision.}
\label{fig:dalitzCutOff}
\end{figure}

\begin{figure}
\begin{center}
\includegraphics[width=0.5\textwidth]{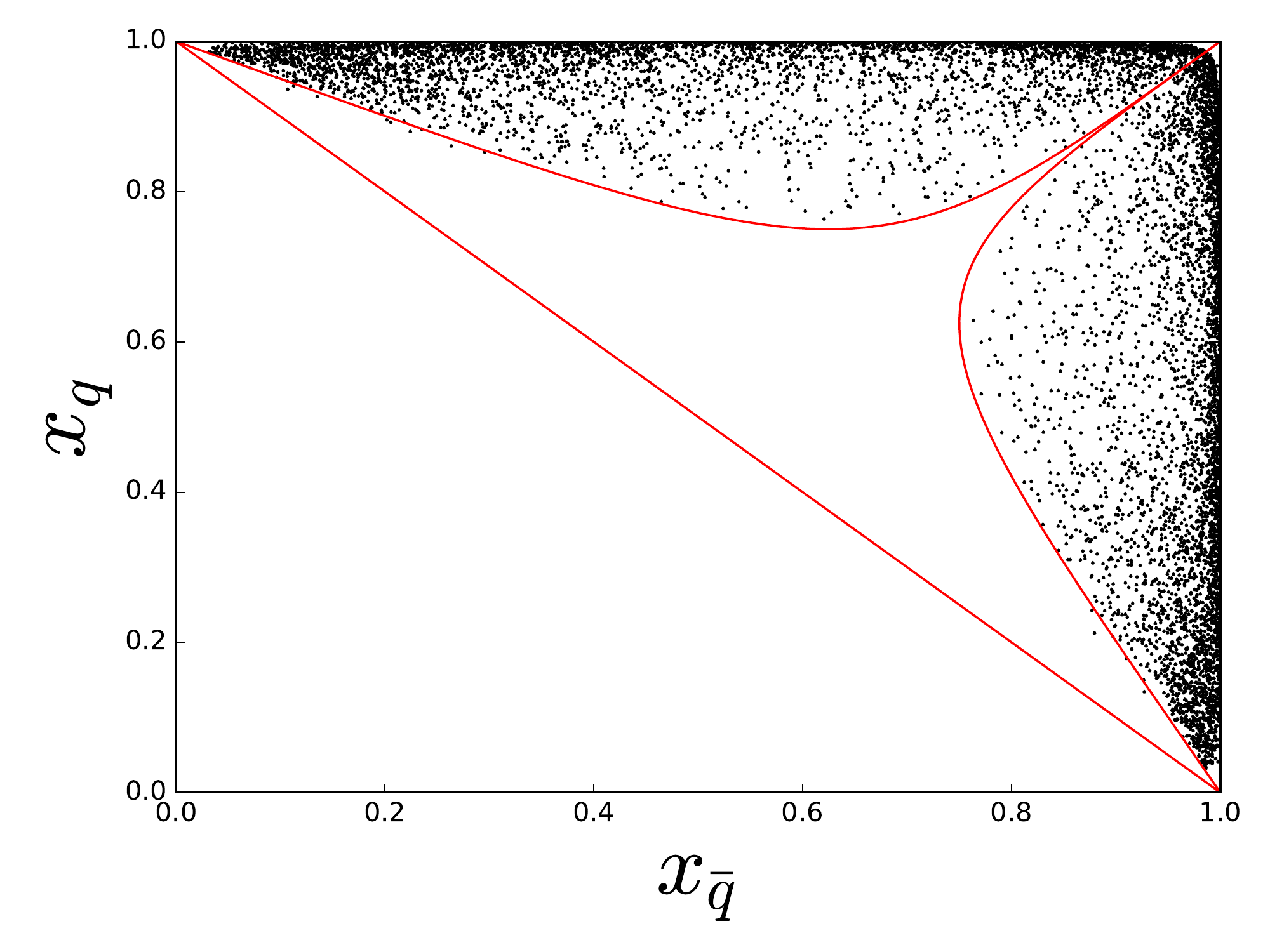}
\end{center}
\caption{Dalitz plot for $e^+e^-\to q \bar q$ showing the region of phase space filled
after multiple emission from the quark and antiquark in the angular-ordered parton shower.
The virtuality of the branchings was preserved in the case of multiple emission.
The line shows the limits for the parton-shower emission for a single emission. $x_i = 2 E_i/Q$ where $E_i$
is the energy of parton $i$ and $Q$ is the centre-of-mass energy of the collision.}
\label{fig:dalitzOffShell}
\end{figure}

The choice of whether to preserve the transverse momentum or virtuality of the branching
affects the phase-space region which is filled by the shower in
the case of multiple emission. 
In this case we cluster the partons
using the Durham jet algorithm \cite{Catani:1991hj}, using the p-scheme as implemented in FastJet~\cite{Cacciari:2011ma},
keeping track of the partons emitted by the quark and antiquark and then take the
hardest additional jet to be the gluon.
The resulting Dalitz plots of $e^+e^-\to q \bar{q}$ show
that while the choice to preserve the transverse momentum of the branching leads to 
a significant number of events in the {\it dead-zone}, Fig.\,\ref{fig:dalitzCutOff},
if the virtuality of the branching is preserved, Fig.\,\ref{fig:dalitzOffShell}, there
is little emission outside the original angular-ordered region.

\section{Hadronization and Colour Reconnection}
\label{sec:hadron}

All the \textsf{Herwig} family of event generator generators use the cluster hadronization model~\cite{Webber:1983if}.
This model is based on the phenomena of colour pre-confinement, i.e. if we non-perturbatively split
the gluons left at the end of the parton shower into quark-antiquark pairs and cluster quarks
and antiquarks into colour-singlet clusters the mass spectrum of these clusters is
peaked at masses close to the cut-off in the parton shower, falls rapidly as the 
cluster mass increases, and is universal, i.e. the mass distribution
of these clusters is independent of the hard scattering process and its centre-of-mass energy.
The cluster model assumes that these clusters are a superposition
of heavy hadronic states and uses a simple phase-space model for their decay into two hadrons.
The main parameters of the model are therefore:
\begin{itemize}
\item the non-perturbative gluon mass, which is not very sensitive and we do not tune;
\item the parameters which control the probability of producing baryons and strange
quarks during cluster decay;
\item the parameter which controls the Gaussian
smearing of the direction of the hadrons produced which contain a parton
from the perturbative evolution about the direction of that parton, 
with separate values for light, charm and bottom quarks.
\end{itemize}
 
There are however a small fraction of large mass clusters for which the two hadron decay ansatz is not reasonable 
and these must first be fissioned into lighter clusters. While only a small fraction of 
clusters undergo fission due to the larger masses of these clusters they produce
a significant fraction of the hadrons.

 A cluster is split into two clusters if the mass, $M$, is such that
\begin{equation}
M^{\bf Cl_{pow}} \geq {\bf Cl_{max}}^{\bf Cl_{pow}}+(m_1+m_2)^{\bf Cl_{pow}},
\label{eqn:clustersplit}
\end{equation}
  where ${\bf Cl_{max}}$ and ${\bf Cl_{pow}}$ are parameters of the model,
  and $m_{1,2}$ are the masses of the constituent partons of the cluster.

\begin{figure*}
\begin{center}
\includegraphics{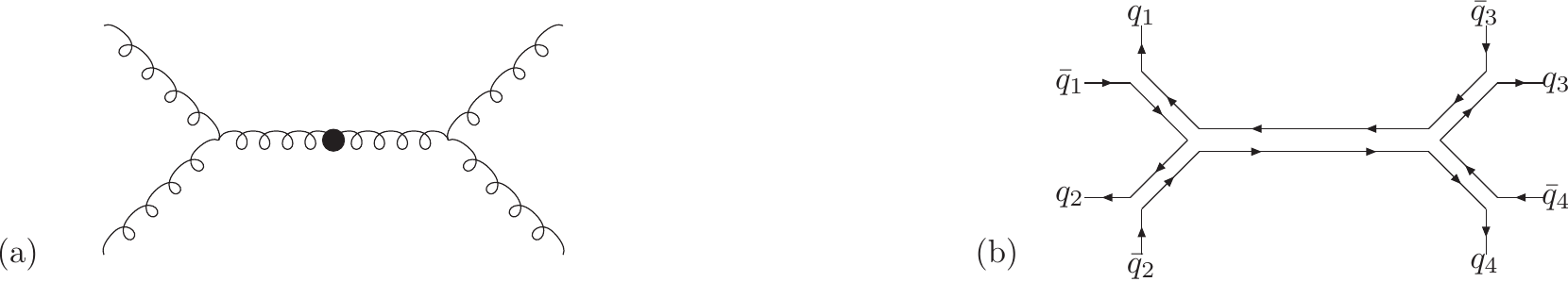}
\end{center}
\caption{Example of colour-singlet gluon pair production followed by the branching of
all the colours via $g\to gg$. The Feynman diagram is shown in (a) whereas the colour flows, including
the non-perturbative splitting of the gluons into quark-antiquark pairs is shown in (b).}
\label{fig:recon}
\end{figure*}
For clusters that need to be split, a $q\bar{q}$ pair is selected to be popped from
the vacuum. The mass distribution of the new clusters is given by
\begin{subequations}
\begin{eqnarray}
M_1 &=& m_1+(M-m_1-m_q)\mathcal{R}_1^{1/P_{\rm split}},\\
M_2 &=& m_2+(M-m_2-m_q)\mathcal{R}_2^{1/P_{\rm split}},
\end{eqnarray}\label{eqn:clusterspect}\end{subequations}
where $m_q$ is the mass of the parton popped from the vacuum, $M_{1,2}$ are
the masses of the clusters formed by the splitting and $\mathcal{R}_{1,2}$ 
are pseudo-random numbers uniformly distributed between 0 and 1. The distribution of the 
masses of the clusters is controlled by the parameter $P_{\rm split}$.

In order to improve the description of charm and bottom hadron production these
parameters for cluster fission all depend on the flavour of the partons in the cluster so that
there are separate parameters for light, charm and bottom quarks.

In practice there is always a small fraction of clusters that are too
light to decay into two hadrons. Before \textsf{Herwig 7.1}
these clusters were decayed to the
lightest hadron, with the appropriate flavours. However in some cases,
for example for clusters containing a charm or bottom quark-antiquark 
pair, or a bottom quark and a light antiquark, there can be a  
number of hadrons of the appropriate flavour below the threshold.
In these cases the lightest meson with the appropriate flavours
is the pseudoscalar $\phantom{1}^1S_0$ state and the vector $\phantom{1}^3S_1$
state is also below the threshold\footnote{For charmonium and bottomonium states there are a number of other states below the
threshold.} which leads to a lower production rate for the vector state with respect to the pseudoscalar state than expected.
For the mesons composed of a bottom quark and a light quark the rate is significantly less than that expected
from the counting of spin states, or indeed observed experimentally~\cite{Acciarri:1994qv,Abreu:1995ky,Buskulic:1995mt,Ackerstaff:1996gz}.
For charmonium and bottomonium states as this mechanism is the only way the vector states
can be produced via hadronization it leads to a complete absence of direct $J/\psi$ and $\Upsilon$ production.
In  \textsf{Herwig 7.1} we therefore include the possibility that instead of just producing the lightest
state all states below the threshold are produced with a probability proportional to $2S+1$, where $S$ is
the spin of the particle.

In order to improve the behaviour at the threshold for charm and bottom clusters 
the option exists of allowing clusters above the threshold mass,
$M_{\rm threshold}$, for the production of two hadrons to decay into a single
  hadron such that a single hadron can be formed for masses
\begin{equation}
M<M_{\rm limit} = (1+{\bf SingleHadronLimit})M_{\rm threshold},
\label{eqn:singlehadron}
\end{equation}
where {\bf SingleHadronLimit} is a free parameter of the model.
  The probability of such a single-meson cluster decay is assumed to 
  decrease linearly for $M_{\rm threshold}<M<M_{\rm limit}$ and there are separate
 parameters for charm and bottom clusters.

 In order to explain the rising trend of $\avrpT$
 (average transverse momentum as a function of the number of charged particles in the event) observed already 
by UA1~\cite{Albajar:1989an} and describe Underlying Event~\cite{Affolder:2001xt,Aad:2010fh,Chatrchyan:2011id,Seymour:2013qka} and the 
Minimum Bias data\cite{Aad:2010ac,Aad:2016mok,Aad:2016xww,Aaboud:2016itf}, 
the hadronization
model is supplemented with a model of colour reconnections (CR)~\cite{Gieseke:2012ft}. The
default version of the model implemented in \textsf{Herwig  7.0} is not very sophisticated. 
The colour reconnection model defines the distance between two partons based on their 
invariant mass, i.e. the distance is small when their invariant mass (cluster mass) 
is small. The aim of the CR model is to reduce the colour length 
$\lambda \equiv \sum_{i=1}^{N_{cl}}m_i^2$, where $N_{cl}$ is the number of clusters 
in an event and $m_i$ is the invariant mass of cluster~$i$. 
The colour reconnection of the clusters leading to a reduction of $\lambda$ 
is accepted with a given probability which is a parameter of the model.
Although the default model is quite simple it should be stressed that its results resemble
the more sophisticated statistical colour reconection model~\cite{Gieseke:2012ft} 
which implements the minimization of $\lambda$ as Metropolis-like algorithm and requires a quick ``cooling`` 
of the random walk.

In this model the only possible reconnections which are not allowed are
connecting the quark and antiquark produced in the non-perturbative splitting of the gluon.
It is therefore possible that the colour lines of a gluon
produced at any other stage of the shower can be reconnected leading to the
production of a colour-singlet object. While this is physically possible
we would expect that it occurs at a rate which is suppressed in the number of
colours, $N_C$, as $\sim\frac1{N_C^2}=\frac19$, not the much higher reconnection
rate $\sim 2/3$ \footnote{The value from the tune of Herwig 7.1 with a new soft and 
diffractive model~\cite{Gieseke:2016fpz}.} which is necessary to describe the underlying event data. 
This can lead to the production of a colour-singlet gluon jet at a much higher
rate than expected. This is particularly problematic in the theoretically clean,
but experimentally inaccessible, colour-singlet gluon pair production
processes often used to study gluon jets~\cite{Gras:2017jty}.

Consider, for example,
the simple process of colour-{\linebreak}singlet gluon pair production followed by the branching of all
the gluons via $g\to gg$, shown in Fig.\,\ref{fig:recon}a. After the non-perturbative
splitting of the gluons into quark-antiquark pairs, as shown in  Fig.\,\ref{fig:recon}b,
without colour reconnection the quarks and antiquarks will be formed
into colour-singlet clusters as $(q_1,\bar{q}_3)$, $(q_3,\bar{q}_4)$, $(q_4,\bar{q}_2)$ and
$(q_2,\bar{q}_1)$. Given the configuration it is likely that the clusters
containing partons from the parton shower of each of the original gluons, i.e. $(q_1,\bar{q}_3)$
and $(q_4,\bar{q}_2)$, will have large masses and the rearrangement to give the
clusters $(q_1,\bar{q}_2)$ and $(q_4,\bar{q}_3)$ will be kinematically favoured, although
it means the original gluons will effectively become colour singlets rather than octets.

In \textsf{Herwig 7.1} we have therefore included the possibility to forbid the colour
reconnection model making any reconnection which would lead to a gluon produced in any stage of
the parton-shower evolution becoming a colour-singlet after hadronization.
We will investigate the effect of this change on the simulation of quark and gluon jets.

\section{Tuning}
\label{sec:tuning}

The \textsf{Rivet}~\cite{Buckley:2010ar} program was used to analyse the simulated events and
compare the results with the experimental measurements. The \textsf{Professor} 
program~\cite{Buckley:2009bj} was then used to interpolate the shower response and
tune the parameters by minimising the chi-squared.\footnote{While tuning the
parameters sensitive to bottom quarks it proved impossible to get a reliable interpolation of the generator response with \textsf{Professor} and therefore a random scan of the bottom parameters
was performed and the values adjusted by hand about the minimum to minimise the $\chi^{\prime2}$.}

In general we use a heuristic chi-squared function
\begin{equation}
  \chi^{\prime2}(\boldmath{p})  = \sum_{\mathcal{O}} w_{\mathcal{O}}  \sum_{b\exists\mathcal{O}}
    \frac{\left(f_b(\boldmath{p})-\mathcal{R}_b\right)^2}{\Delta_b^2}
\end{equation}
where $\boldmath{p}$ is the set of parameters being tuned, $\mathcal{O}$ are the observables
used each with weight $w_{\mathcal{O}}$, $b$ are the different bins in each observable distribution
with associated experimental measurement $\mathcal{R}_b$ , error $\Delta_b$ and Monte Carlo
prediction $f_b(\boldmath{p})$. Weighting of those observables for which a good description of
the experimental result is important is used in most cases. The parameterisation
of the event generator response, $f(\boldmath{p})$, is used to minimize $\chi^{\prime2}$ and find
the optimum parameter values. We take $w_{\mathcal{O}}=1$ in most cases except for the particle multiplicities where
we use $w_{\mathcal{O}}=10$ and total charged particle multiplicities where we use $w_{\mathcal{O}}=50$.
This ensures that particle multiplicities influence the result of the fit and are required due to the much higher
quantity of event shape and spectrum data used in the tuning.
Given the
aim of this paper is to improve the description of gluon jets this data was also included with $w_{\mathcal{O}}=10$
in order to avoid the fit being dominated by the large quantity of data sensitive to quark jets.
In addition as we do not except a Monte Carlo event generator to give a perfect description of all the data and
in order to avoid the fit being dominated by a few observables with very small experimental errors we use
\begin{equation}
\Delta^{\rm eff}_b = \max(0.05\times\mathcal{R}_b,\Delta_b),   
\end{equation}
rather than the true experimental error, $\Delta_b$, in the fit.

The standard procedure which was adopted to tune the shower
and hadronization parameters of the \textsf{Herwig++} and 
\textsf{Herwig\ 7} event generators to data is
\begin{itemize}
\item first the shower and those
hadronization parameters\linebreak which are primarily sensitive to light quark-initiated processes
are tuned to LEP1 and SLD measurements of event shapes, the average charged multiplicity and\linebreak
charged multiplicity distribution, and identified particle
spectra and rates which only involve light quark mesons and baryons;
\item the hadronization parameters for bottom quarks are\linebreak tuned to
the bottom quark fragmentation function measured by LEP1 and SLD together with
LEP1 and SLD measurements of event shapes and identified particle\linebreak  spectra from bottom events;
\item the hadronization parameters involving charm quarks are then tuned
to identified particle spectra, from both the B-factories and LEP1,
and LEP1 and SLD measurements of event shapes and identified particle spectra from charm events;
\item the light quark parameters are then retuned using the new values of the
  bottom and charm parameters together with different weights for the
  charged multiplicity distributions in $e^+e^-$ collisions at energies between 12\,GeV and 209\,GeV
  due to the difficulty in fitting the charged multiplicity.
\end{itemize}
Only $e^+e^-$ annihilation data from the continuum region near the $\Upsilon(4s)$ meson, for charm meson spectra,
and at the Z-pole from LEP1 and SLD were used in the tune.

In this paper we have extended this approach in order to better constrain the energy evolution
to include data from a wider range of centre-of-mass energies both below the Z-pole, from the
JADE and TASSO experiments, and above the Z-pole, from LEP2.

In order to tune the shower and light quark hadronization parameters we used
data on jet rates and event shapes for centre-of-mass energies between 14 and 44\,GeV~\cite{Braunschweig:1990yd,MovillaFernandez:1997fr,Pfeifenschneider:1999rz}, 
at LEP1 and SLD~\cite{Abreu:1996na,Barate:1996fi,Pfeifenschneider:1999rz,Abbiendi:2004qz,Heister:2003aj} and LEP2~\cite{Pfeifenschneider:1999rz,Heister:2003aj,Abbiendi:2004qz}, 
particle multiplicities~\cite{Abreu:1996na,Barate:1996fi} and spectra~\cite{Akers:1994ez,Alexander:1995gq,Alexander:1995qb,Abreu:1995qx,Alexander:1996qj,Abreu:1996na,Barate:1996fi,Ackerstaff:1997kj,Abreu:1998nn,Ackerstaff:1998ap,Ackerstaff:1998ue,Abbiendi:2000cv,Heister:2001kp} at LEP 1, identified particle spectra
below the $\Upsilon(4S)$ from Babar~\cite{Lees:2013rqd}, 
the charged particle multiplicity~\cite{Ackerstaff:1998hz,Abe:1996zi} and particle spectra~\cite{Ackerstaff:1998hz,Abe:1998zs,Abe:2003iy} in light quark events at LEP1 and SLD,
the charged particle multiplicity in light quark events at\linebreak LEP2~\cite{Abreu:2000nt,Abbiendi:2002vn}, the charged particle multiplicity distribution at LEP 1\cite{Decamp:1991uz},
and hadron multiplicities at the Z-pole \cite{Amsler:2008zzb}.
We also implemented in Rivet and made use of the data on the properties of gluon 
jets~\cite{Abbiendi:2003gh,Abbiendi:2004pr} for the first time.

\begin{table*}
  \begin{tabular}{|c|ccc|ccc|ccc|ccc|}
    \hline
    Cut-Off    & \multicolumn{6}{c|}{$p_\perp$} & \multicolumn{6}{c|}{Virtual Mass}\\
    \hline
    Preserved  & \multicolumn{3}{c|}{$p_\perp$}& \multicolumn{3}{c|}{$q^2$}
               & \multicolumn{3}{c|}{$p_\perp$}& \multicolumn{3}{c|}{$q^2$}\\
    Tune       & A & B & C& A & B & C& A & B & C& A & B & C\\
\hline
\multicolumn{13}{|c|}{Bottom quark hadronization parameters}\\
\hline
ClMaxBottom             & \multicolumn{3}{c|}{4.655}  & \multicolumn{3}{c|}{3.911} & \multicolumn{3}{c|}{4.0612} & \multicolumn{3}{c|}{4.163} \\
ClPowBottom             & \multicolumn{3}{c|}{0.622}  & \multicolumn{3}{c|}{0.638 } & \multicolumn{3}{c|}{0.9475} & \multicolumn{3}{c|}{0.590} \\
PSplitBottom            & \multicolumn{3}{c|}{0.499}  & \multicolumn{3}{c|}{0.531  } & \multicolumn{3}{c|}{1.9568} & \multicolumn{3}{c|}{1.881} \\
ClSmrBottom             & \multicolumn{3}{c|}{0.082}  & \multicolumn{3}{c|}{0.020 } & \multicolumn{3}{c|}{0.04  } & \multicolumn{3}{c|}{0.040   } \\
SingleHadronLimitBottom & \multicolumn{3}{c|}{0.000    }  & \multicolumn{3}{c|}{ 0.000   } & \multicolumn{3}{c|}{0.0204} & \multicolumn{3}{c|}{0.000    } \\
\hline
\multicolumn{13}{|c|}{Charm quark hadronization parameters}\\
\hline
SingleHadronLimitCharm  & \multicolumn{3}{c|}{0.000   }  & \multicolumn{3}{c|}{0.000} & \multicolumn{3}{c|}{0.078}  & \multicolumn{3}{c|}{0.012 } \\
ClMaxCharm              & \multicolumn{3}{c|}{3.551}  & \multicolumn{3}{c|}{3.638} & \multicolumn{3}{c|}{3.805}  & \multicolumn{3}{c|}{3.885 } \\
ClPowCharm              & \multicolumn{3}{c|}{1.923}  & \multicolumn{3}{c|}{2.332} & \multicolumn{3}{c|}{2.242}  & \multicolumn{3}{c|}{2.452 } \\
PSplitCharm             & \multicolumn{3}{c|}{1.260}  & \multicolumn{3}{c|}{1.234} & \multicolumn{3}{c|}{1.895}  & \multicolumn{3}{c|}{1.767 } \\
ClSmrCharm              & \multicolumn{3}{c|}{0.000   }  & \multicolumn{3}{c|}{0.000   } & \multicolumn{3}{c|}{0.000}  & \multicolumn{3}{c|}{0.000 } \\
\hline
\multicolumn{13}{|c|}{Light quark hadronization and shower parameters}\\
\hline
AlphaMZ ($\alpha^{\rm CMW}_S(M_Z)$)       & 0.1094 & 0.1087 & 0.1126 & 0.1260 & 0.1262 & 0.1265 & 0.1221 & 0.1218 & 0.1184 & 0.1314 & 0.1317 & 0.1254\\
pTmin          & 1.037  & 0.933 & 0.809 & 1.301  & 1.223  & 0.992  & \multicolumn{3}{c|}{N/A}  & \multicolumn{3}{c|}{N/A}  \\
aParameter              & \multicolumn{3}{c|}{N/A}  & \multicolumn{3}{c|}{N/A} & \multicolumn{3}{c|}{0.367} & \multicolumn{3}{c|}{0.234} \\
cutoffKinScale & \multicolumn{3}{c|}{N/A}  & \multicolumn{3}{c|}{N/A} & 2.939   & 2.910   & 2.294   & 3.277   & 3.279    & 1.938 \\
ClMaxLight      & 3.504 & 3.639 & 4.349 & 3.058  & 3.003 & 3.197 & 3.328  & 3.377  & 3.846  & 3.414  & 3.427  & 3.477 \\
ClPowLight      & 2.576 & 2.575 & 1.226 & 1.513  & 1.424 & 2.786 & 1.286  & 1.318  & 2.063  & 2.766  & 2.792  & 2.35  \\
PSplitLight     & 1.003 & 1.016 & 0.855 & 0.885  & 0.848 & 0.648 & 1.198  & 1.185  & 1.277  & 1.346  & 1.333  & 2.015 \\
PwtSquark       & 0.552 & 0.597 & 1.167 & 0.602  & 0.666 & 1.024 & 0.721 & 0.741 & 0.782  & 0.626  & 0.646 & 1.15  \\
PwtDIquark      & 0.369 & 0.344 & 0.181 & 0.416  & 0.439 & 0.512 & 0.277 & 0.273  & 0.246 & 0.321 & 0.328 & 0.366\\
\hline
  \end{tabular}
  \caption{The Monte Carlo parameters obtained for different choices of the cut-off option, the preserved quantity in the 
  shower and weight of the charged particle multiplicity data.}
  \label{tab:parameters}
\end{table*}
\begin{table*}
  \begin{center}
\begin{tabular}{|c|ccc|ccc|ccc|ccc|c|}
    \hline
    Cut-Off    & \multicolumn{6}{c|}{$p_\perp$} & \multicolumn{6}{c|}{Virtual Mass} & Number of \\
    \cline{1-13}
    Preserved  & \multicolumn{3}{c|}{$p_\perp$}& \multicolumn{3}{c|}{$q^2$}
               & \multicolumn{3}{c|}{$p_\perp$}& \multicolumn{3}{c|}{$q^2$} & degrees of\\
    Tune       & A & B & C& A & B & C& A & B & C& A & B & C & freedom\\
    \cline{1-13}
\multicolumn{13}{|c|}{Tuning Observables}&(sum including weights)\\
\hline
Light quarks & 4.4 & 4.3 & 6.7 & 3.0 & 2.9 & 4.2 & 7.8 & 7.6 & 6.9 & 4.6 & 4.3 & 3.6 & 10122(14099)\\
Charm  quarks & 3.2 & 2.8 & 5.8 & 3.6 & 3.5 & 3.9 & 4.5 & 4.6 & 6.4 & 3.9 & 3.9 & 7.4 &549(891)\\
Bottom quarks & 4.0 & 3.4 & 3.6 & 5.4 & 4.9 & 3.4 & 3.4 & 3.3 & 3.4 & 4.1 & 4.1 & 4.9 &346(1309)\\
Gluons & 1.1 & 1.1 & 1.5 & 1.1 & 1.1 & 1.4 & 1.2 & 1.2 & 1.2 & 1.3 & 1.2 & 1.5 &188(1880) \\
\hline
\multicolumn{13}{|c|}{$N_{\rm charged}$}&\\
\hline
Gluon & 14.2 & 18.6 & 22.6 & 26.9 & 37.1 & 60.0 & 3.4 & 3.7 & 8.1& 10.0 & 11.0 & 22.8 &26\\
All quarks & 4.6 & 2.7 & 2.7 & 3.4 & 2.5 & 5.2 & 11.6 & 10.7 & 3.7 & 7.2 & 6.5 & 1.6 &48\\
Light quarks & 2.2 & 1.7 & 2.8 & 1.7 & 1.8 & 4.4 & 4.8 & 4.4 & 2.1 & 3.9 & 3.5 & 1.8 &27\\
Charm quarks & 2.8 & 2.0 & 1.1 & 2.2 & 1.6 & 1.0 & 2.8 & 2.6 & 1.2 & 2.2 & 2.2 & 0.9 &17\\
Bottom quarks & 20.4 & 18.1 & 15.8 & 24.1 & 21.3 & 15.7 & 33.4 & 33.1 & 34.7 & 22.0 & 21.5 & 46.2 &27\\
ATLAS Jets & 3.2 &0.9&4.3&13.3&10.1&7.8&21.8&19.0&6.4&33.3&31.3&38.0&22\\ 
\hline
\end{tabular}
  \end{center}
  \caption{The values of $\chi^2$ per degree of freedom obtained in the fit for different choices of the cut-off option, 
  the preserved quantity in the shower and weight of the charged particle multiplicity data. The values are $\chi^{\prime2}$
  as described in the text for the tuning observables, normalised to the sum of the weights for the different bins, 
  and the true $\chi^2$ using the experimental error for the charged particle multiplicities. 
  The number of degrees of freedom for each set of observables is given together
  with the sum including weights in brackets, where this is different.}
  \label{tab:chisq}
\end{table*}

The hadronization parameters for charm quarks were\linebreak tuned using
the charged multiplicity in charm events at\linebreak SLD~\cite{Abe:1996zi} and
LEP2~\cite{Abreu:2000nt,Abbiendi:2002vn}, the light hadron spectra in charm events at LEP1 and SLD \cite{Ackerstaff:1998hz,Abe:1998zs,Abe:2003iy},
the multiplicities of charm hadrons at the Z-pole \cite{Abreu:1996na,Amsler:2008zzb}, and
charm hadron spectra below the $\Upsilon(4S)$~\cite{Seuster:2005tr,Aubert:2006cp} and at LEP1~\cite{Barate:1999bg}.

The hadronization parameters for bottom quarks were tuned using
the charged multiplicity in bottom events at\linebreak SLD~\cite{Abe:1996zi} and
LEP2~\cite{Abreu:2000nt,Abbiendi:2002vn}, the light hadron spectra in bottom events at LEP1 and SLD \cite{Ackerstaff:1998hz,Abe:1998zs,Abe:2003iy},
the multiplicities of charm and bottom hadrons at the Z-pole \cite{Abreu:1996na,Amsler:2008zzb},
charm hadron spectra at LEP1~\cite{Barate:1999bg} and the
bottom fragmentation function measured at LEP1 and SLD~\cite{Abe:2002iq,Heister:2001jg,DELPHI:2011aa}.

In order to tune the evolution of the total charged particle multiplicity in $e^+e^-$ collisions as a function of energy the results of Refs.\,\cite{Derrick:1986jx,Aihara:1986mv,Berger:1980zb,Bartel:1983qp,Braunschweig:1989bp,Zheng:1990iq,Acton:1991aa,Abe:1996zi,Abreu:1996na,Abreu:2000nt,Abbiendi:2002vn,Heister:2003aj} spanning energies from 12 to 209 GeV were used.

In order to study the various effects we have discussed we have produced
tunes for the shower and hadronization parameters in the case that either
the transverse momentum or virtuality in the shower is preserved. In each case we first tuned the shower and
light quark parameters without the data on charged particle multiplicities as centre-of-mass energies below the mass of
the $Z^0$ boson. In the final stage of the process where we retune these parameters three tunes were produced for
each choice of cut-off and preserved quantity, one (labelled A) without the low-energy charged multiplicity data, one
(labelled B) where all the\linebreak charged multiplicity data was included with in the tune with weight $w_{\mathcal{O}}=100$ and a final tune
(labelled C) where this data had weight $w_{\mathcal{O}}=1000$.

\begin{figure*}
\begin{center}
\includegraphics[width=0.45\textwidth]{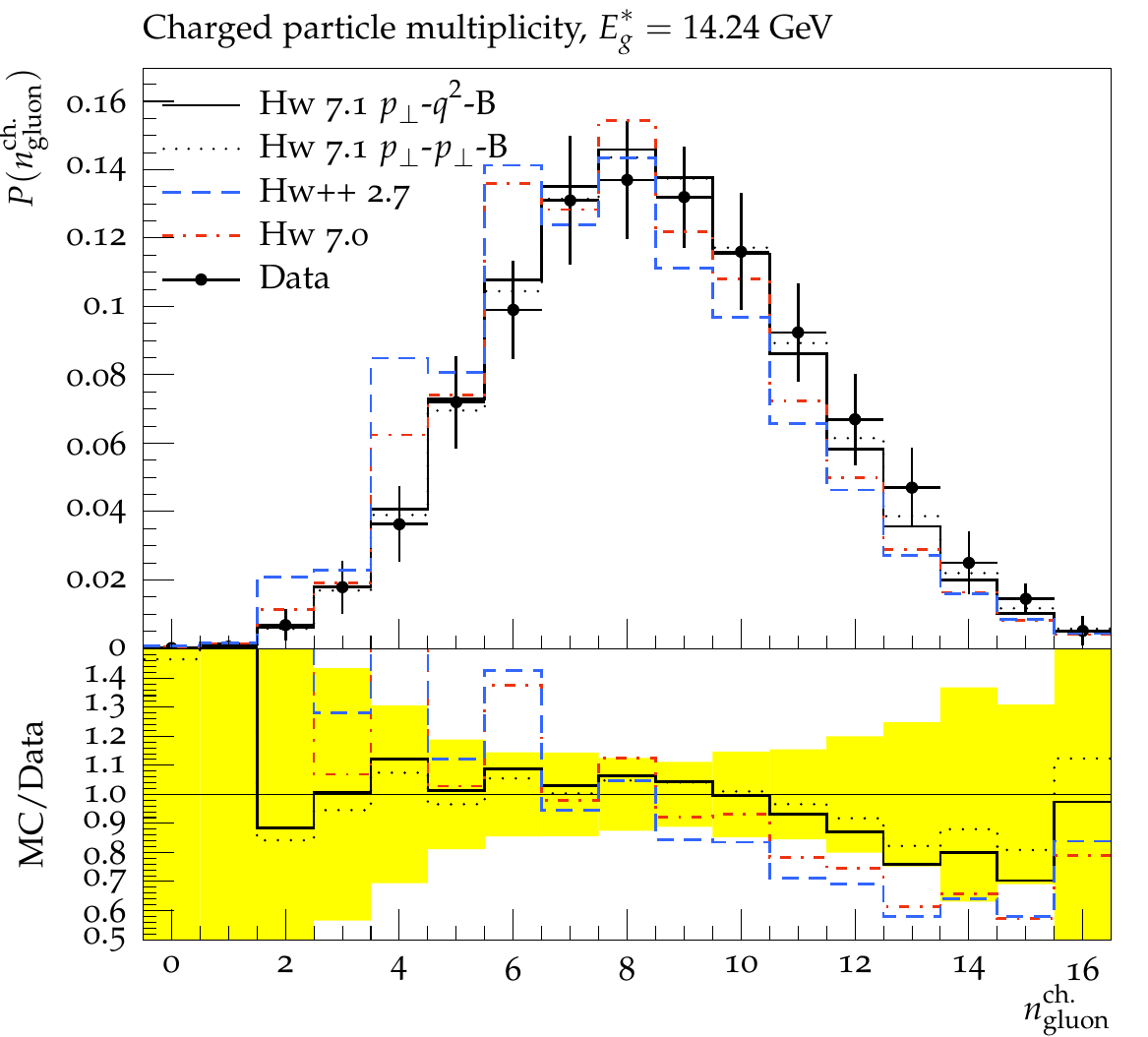}
\includegraphics[width=0.45\textwidth]{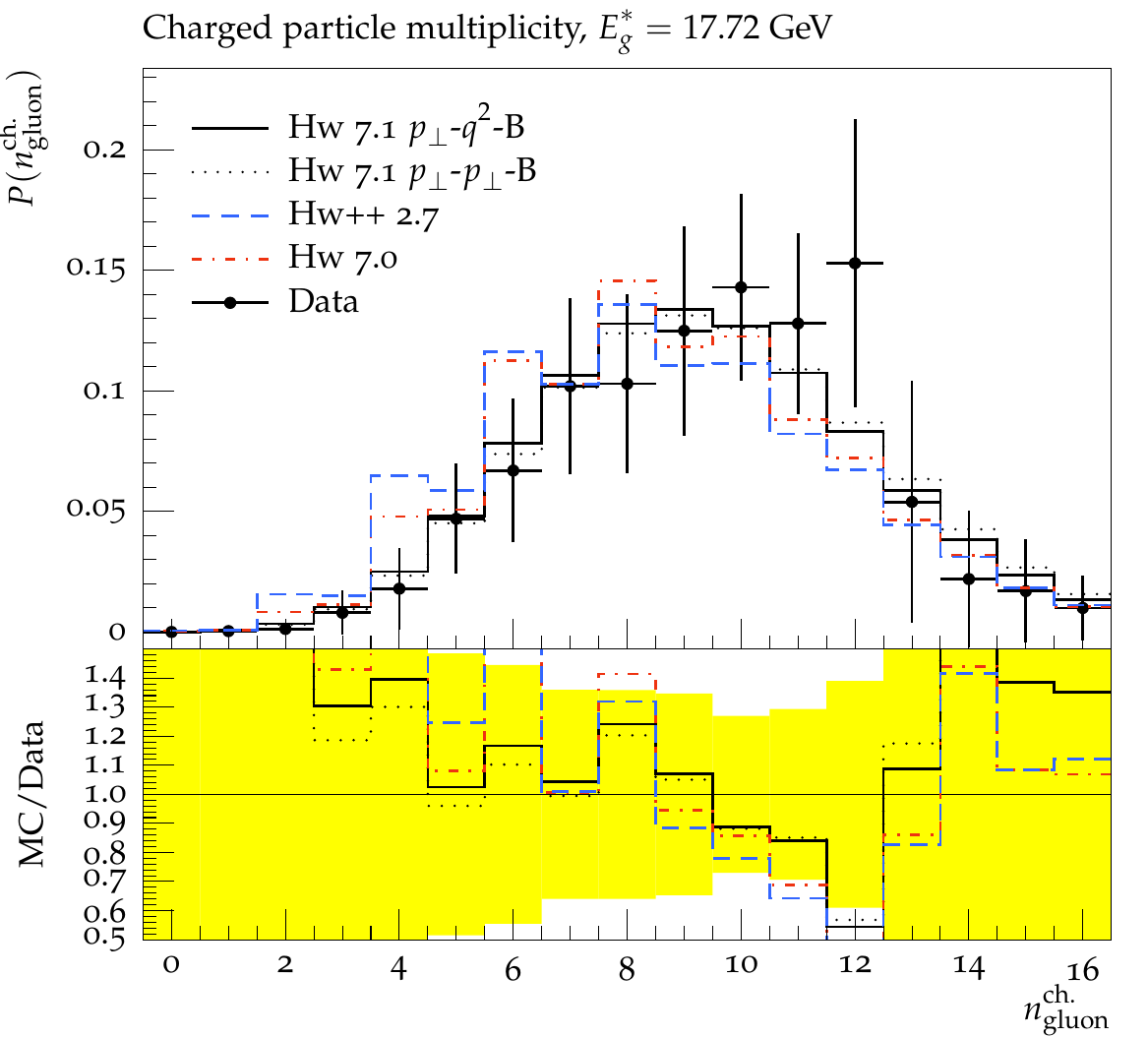}
\end{center}
\caption{Multiplicity distribution of charged particles in gluons jets for two different gluon energies compared to data from OPAL \cite{Abbiendi:2003gh}.}
\label{fig:OPALgluonNcharged}
\end{figure*}

Unfortunately due to the CPU time required it is impossible to include
the ATLAS data~\cite{Aad:2016oit} directly in the tune, therefore we compare
the results of the different tunes to this data.

\section{Results}
\label{sec:results}

We have produced 12 tunes for different choices of the cut-off variable in the shower,
the choice of which quantity to preserve in the parton shower, and different weightings of the
charged particle multiplicities. The parameters obtained in the fits are given
in Table.\,\ref{tab:parameters} while the $\chi^2$ values are given in Table.\,\ref{tab:chisq}.

The effects of changing the colour reconnection model can be seen in Fig.\,\ref{fig:OPALgluonNcharged}. In the results of \textsf{Herwig++ 2.7.1} or
\textsf{Herwig 7.0} there is an unphysical tendency of the gluon jets to contain an even number of charged particles due to the production of colour-singlet gluons by the reconnection model, this feature is not present in any of the new tunes which provide a much better 
description of the distribution of charged particles in the gluon jets, see also the Appendix.

The choice of which tune and choice of cut-off variable and preserved quantity has to be a balance between how well we wish to describe the various different data sets, as unfortunately no choice provides a good description of all the data sets.

\begin{figure}
  \includegraphics[width=0.5\textwidth]{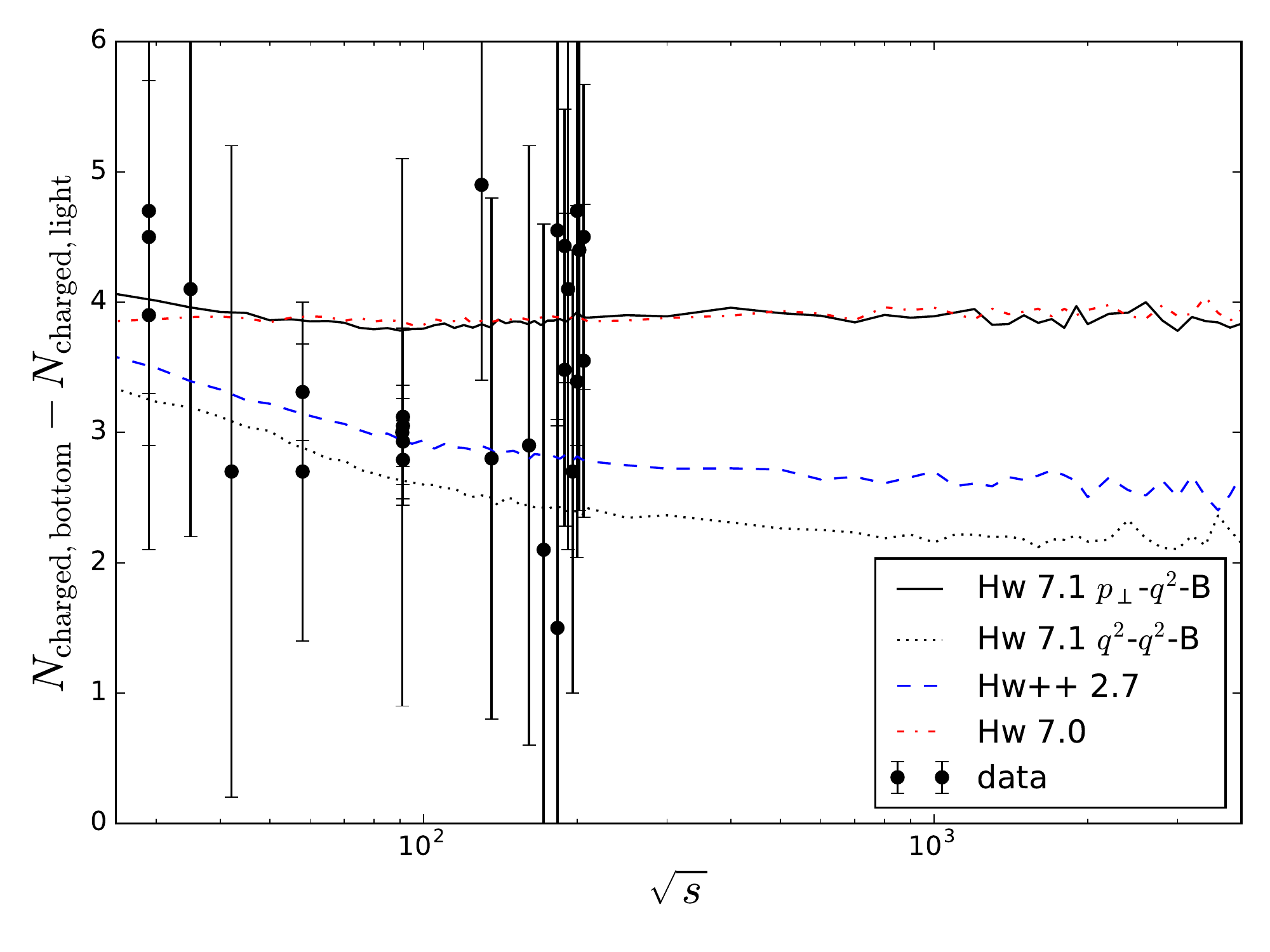}
  \caption{Difference between the charged multiplicity in bottom and light quark events in $e^+e^-$ collisions as a function of centre-of-mass energy.
  The data is from \cite{Rowson:1985xh,Sakuda:1985jd,Aihara:1986mv,Braunschweig:1988gy,Althoff:1983yv,Nagai:1992yd,Okabe:1997uf,Schumm:1992pv,Abreu:1995pk,Akers:1995ww,Abe:1996zi,Abreu:2000nt,Delphi:2002,Abbiendi:2002vn,Abe:2003iy} as compiled in \cite{Dokshitzer:2005ri}}
  \label{fig:bottomdiff}
\end{figure}

If we first consider the choice of cut-off it is clear that using a virtual mass provides a larger $\chi^{\prime2}$
for all sets of observables used in the tuning apart from those sensitive to bottom quarks.
In addition it displays an unphysical energy dependence in the difference in
charged particle multiplicities between bottom~(or charm) quark and light quark events, as shown in Fig.\,\ref{fig:bottomdiff}
where the results which use a cut-off on the virtual mass, \textsf{Herwig++ 2.7.1} and the new tune\linebreak $q^2$-$q^2$-B, 
show a strong dependence on the centre-of-mass energy while those which use a $p_\perp$ cut-off,
\textsf{Herwig 7.0} and the new tune
$p_\perp$-$q^2$-B,  are relatively independent of energy. We therefore prefer a cut-off on the minimum transverse momentum of the branching.

\begin{figure}
\begin{center}
\includegraphics[width=0.5\textwidth]{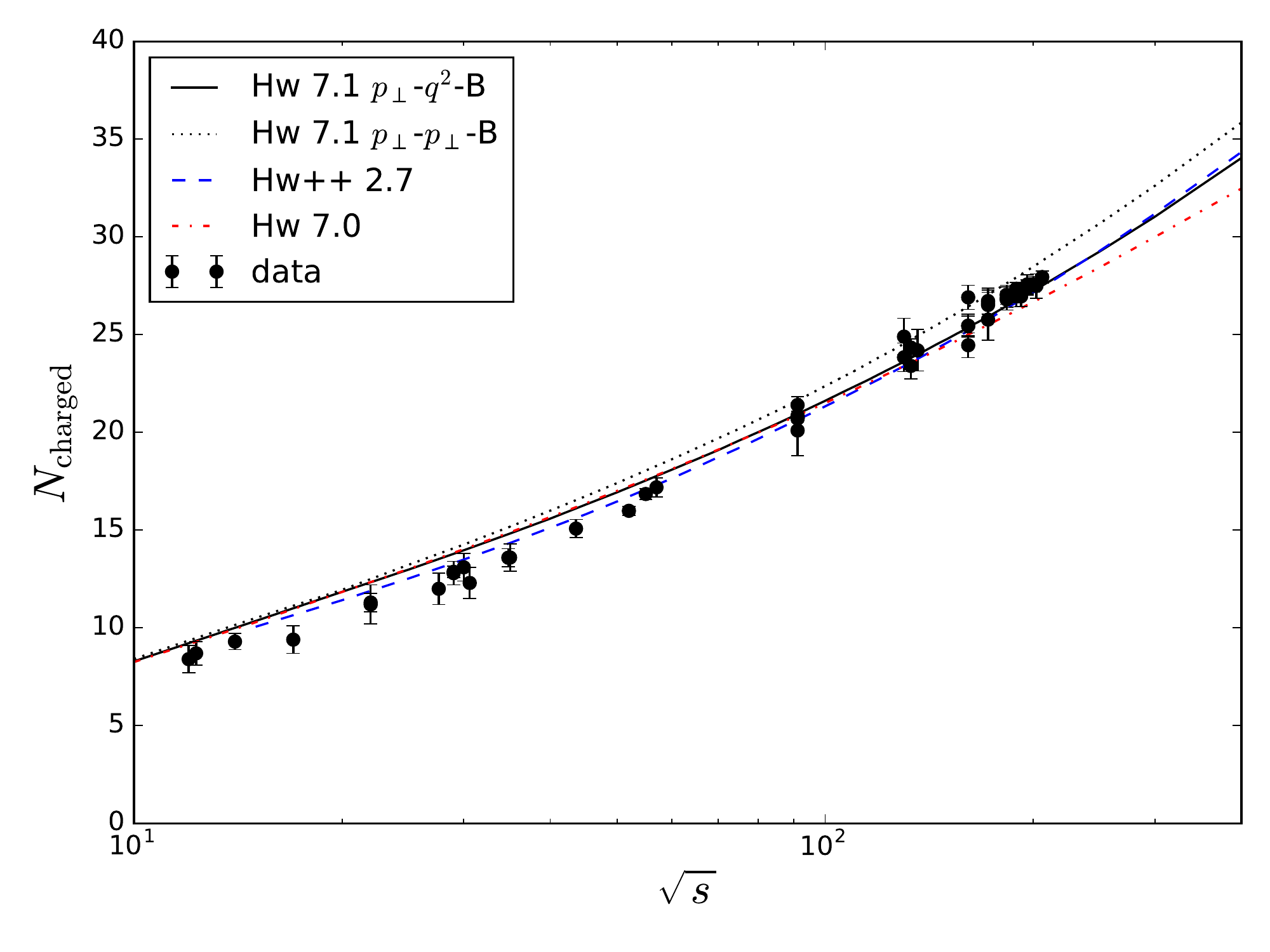}
\end{center}
\caption{The evolution of the number of charged particles in \mbox{$e^+e^-\to{\rm hadrons}$}
         as a function of the centre-of-mass energy.}
\label{fig:allN}
\end{figure}

In order to obtain a reasonable evolution of the number of charged particles with centre-of-mass energy in $e^+e^-$ collisions, see Fig.\,\ref{fig:allN}, without ruining the description of
particle spectra and event shape observables we choose to use the B tune as our default.

\begin{figure}
  \includegraphics[width=0.5\textwidth]{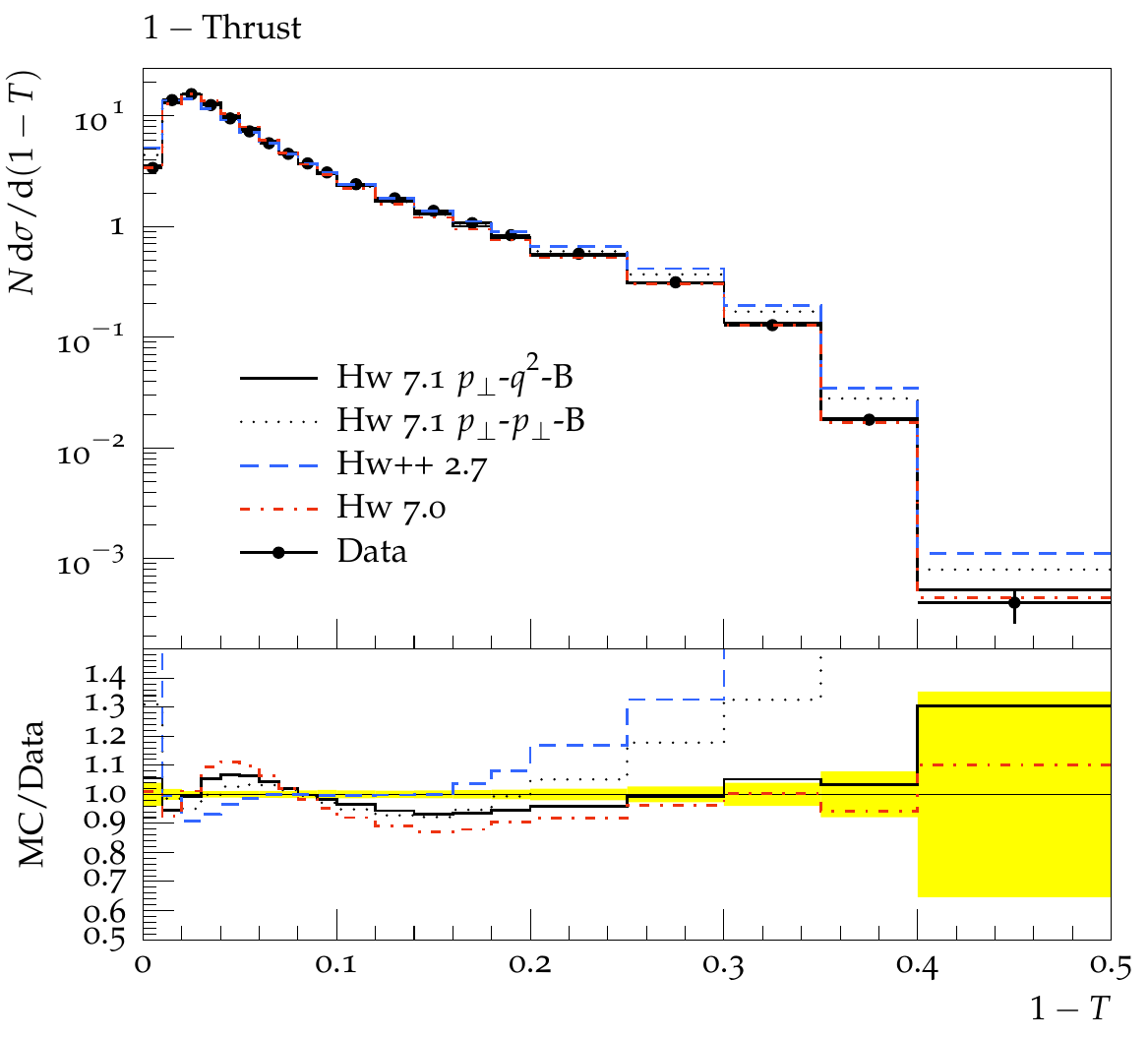}
  \caption{The thrust at the Z-pole compared to data from the DELPHI~\cite{Abreu:1996na} experiment.}
  \label{fig:thrust}
\end{figure}

\begin{figure}
\begin{center}
\includegraphics[width=0.5\textwidth]{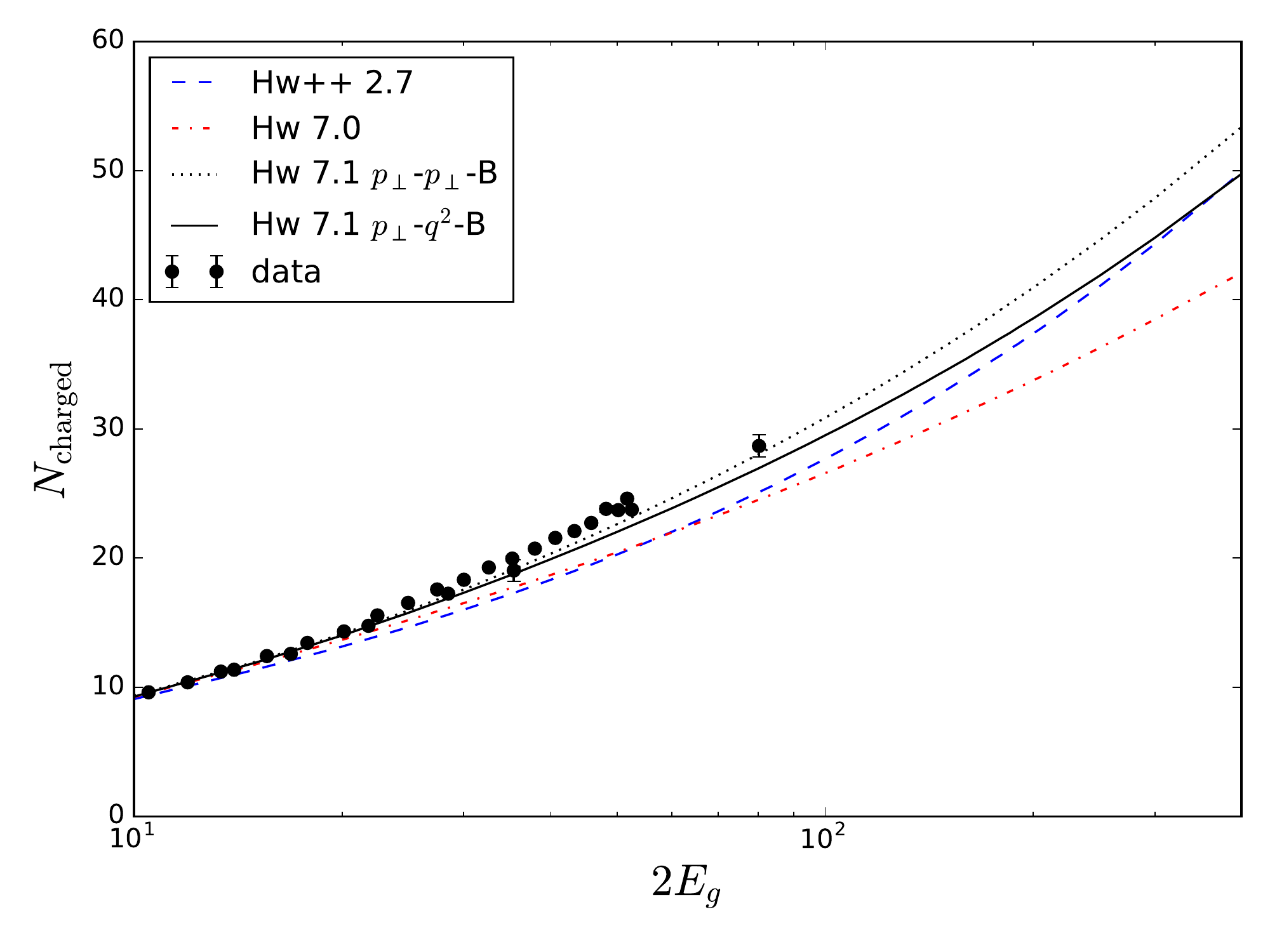}
\end{center}
\caption{The evolution of the number of charged particles in gluon
         jets as a function of twice the energy of the gluon jet.}
\label{fig:gluonN}
\end{figure}

\begin{figure}
  \includegraphics[width=0.5\textwidth]{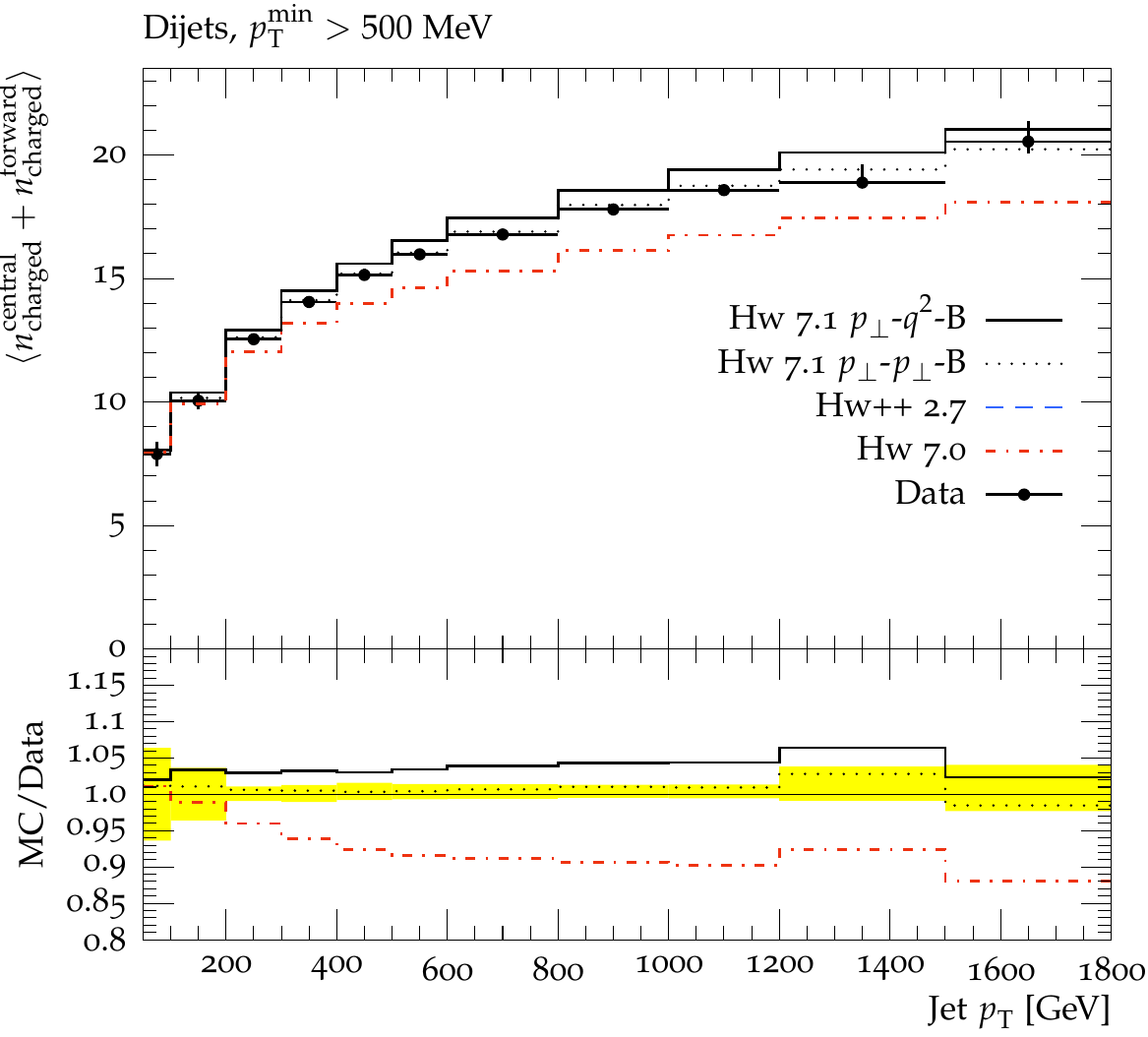}
  \caption{The average number of charged particles in jets as a function of the jet transverse momentum
           compared to data from the ATLAS experiment~\cite{Aad:2016oit}.}
  \label{fig:ATLASNCharged}
\end{figure}

\begin{figure}
  \includegraphics[width=0.5\textwidth]{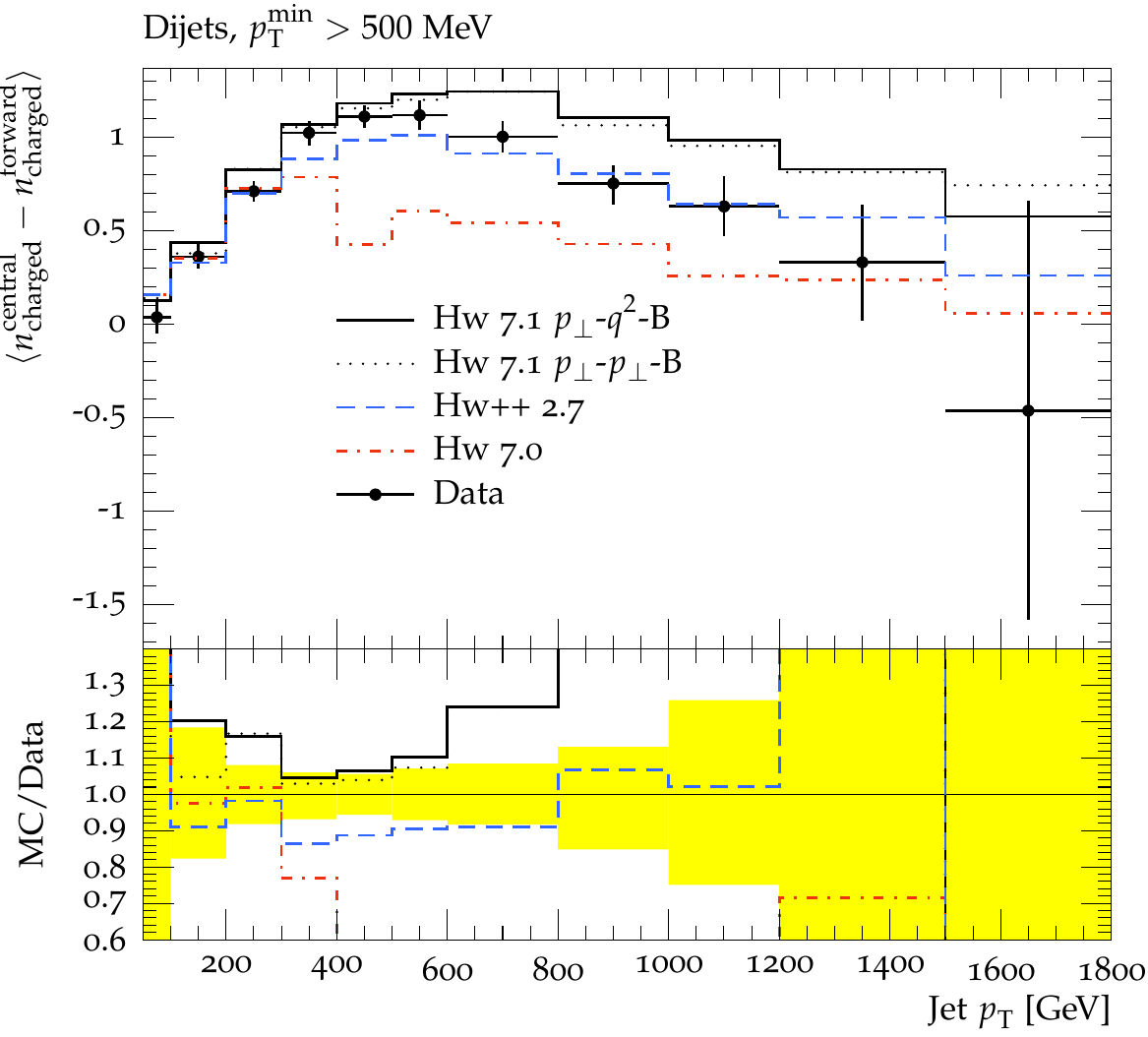}
  \caption{The difference between the average number of particles in central and forward
           jets compared to data from the ATLAS experiment~\cite{Aad:2016oit}.}
  \label{fig:ATLASNdiff}
\end{figure}

The choice of whether to preserve the $p_\perp$ or $q^2$ of the branching is more complicated. While the data on light quark jets,
in particular event shapes measured at LEP (for example the thrust Fig.\,\ref{fig:thrust}), favour preserving $q^2$ the data on the
charged particle multiplicity
in gluon jets at LEP~Fig.\,\ref{fig:gluonN}, and in jets at the LHC~Figs.\,\ref{fig:ATLASNCharged},\ref{fig:ATLASNdiff} favours preserving
the $p_\perp$ of the branching.

Our preferred choice, in particular in the presence of higher-order matching, is to preserve the $q^2$ of the branching in order to
ensure that the parton shower does not overpopulate the dead-zone. This also ensures a more reasonable value of strong coupling,
$\alpha^{\rm CMW}_S(M_Z)=0.126$ which gives $\alpha^{\rm \overline{MS}}_S(M_Z)=0.118$. However given the better description of gluon jets
it is reasonable to also consider the alternative of preserving the $p_\perp$, see for example Fig.~\ref{fig:summary_hadron_pp_all}
from the Appendix.

\section{Conclusions}

We have performed a tuning the the \textsf{Herwig 7} event generator using data on gluon jets from LEP
for the first time. Together with changes to the non-perturbative modelling this gives a significantly better
description of gluon jets, in particular their charge particle multiplicity. It is however impossible
to get a good description of the LEP particle spectra and the charged particle multiplicities, particularly in gluon jets,
at the same time. We therefore choose the tune $p_\perp$-$q^2$-B as the default for \textsf{Herwig~7.1}. However for
jets at the LHC the tune $p_\perp$-$p_\perp$-B gives a better description of jet properties.

While the tunes presented in this paper are an improvement on their predecessors there is a tension between the
data on charged particle multiplicities, for both quark and gluon initiated jets, and the data on event shapes
and particle spectra from LEP. The cluster hadronization model also continues to have problems describing final states
in events with bottom quarks. Any further improvement in the description of this data will require improvements to
the non-perturbative modelling.

\begin{acknowledgements}

This work was supported in part by the European Union as part of the FP7 and
H2020 Marie Sk\l odowska-Curie Initial Training Networks MCnetITN and MCnetITN3
(PITN-GA-2012-315877 and 722104). Daniel Reichelt thanks CERN for
the award of a summer studentship during which this work was initiated and
acknowledges support from the German Research Foundation (DFG) under 
grant No.\ SI 2009/1-1. Andrzej Siodmok acknowledges support from the National
Science Centre, Poland Grant No.~2016/23/D/ST2/02605.
We thank our collaborators on \textsf{Herwig} for many useful discussions.
The tuning of \textsf{Herwig} to experimental data would not have been possible
without the use of GRIDPP computer resources.
\end{acknowledgements}

\appendix

\section{Generalized angularities and quark and gluon jet discrimination power}
\label{appendix}
In this appendix we investigate how the improvements of the simulation of quark and gluon 
proposed in the manuscript affect the quark and gluon jet discrimination power recently 
studied in~\cite{Gras:2017jty}\footnote{The results and the analysis code used for this study is available as a \textsc{Rivet} routine \cite{Buckley:2010ar}, 
which can be downloaded from \url{https://github.com/gsoyez/lh2015-qg}.}. 
For this purpose, we present results for five generalized angularities $\lambda^{\kappa}_{\beta}$~\cite{Larkoski:2014pca}:
\begin{equation*}
\arraycolsep=5pt
\begin{array}{cccccc}
\label{eq:ang}
 (\kappa,\beta)&(0,0) & (2,0) & (1,0.5) & (1,1) & (1,2) \\
\lambda^{\kappa}_{\beta}: & \text{multiplicity} &  p_T^D &  \text{LHA} & \text{width} & \text{mass}
\end{array}
\end{equation*}
where 
$\lambda^{\kappa}_{\beta} = \sum_{i \in \text{jet}} z_i^\kappa \theta_i^\beta,$
 $i$ runs over the jet constituents, $z_i \in [0,1]$ is a momentum fraction, 
and $\theta_i \in [0,1]$ is an angle to the jet axis.
To quantify discrimination performance, we use classifier separation:
\begin{equation*}
\Delta =  \frac{1}{2} \int \text{d} \lambda \, \frac{\bigl(p_q(\lambda) - p_g(\lambda)\bigr)^2}{p_q(\lambda) + p_g(\lambda)},
\end{equation*}
where $p_q$ ($p_g$) is the probability distribution for $\lambda$ in a generated quark jet (gluon jet) 
sample. $\Delta = 0$ corresponds to no discrimination power and $\Delta = 1$ corresponds to perfect 
discrimination power.

\begin{figure}
\centering
{
\includegraphics[width = 1.0\columnwidth]{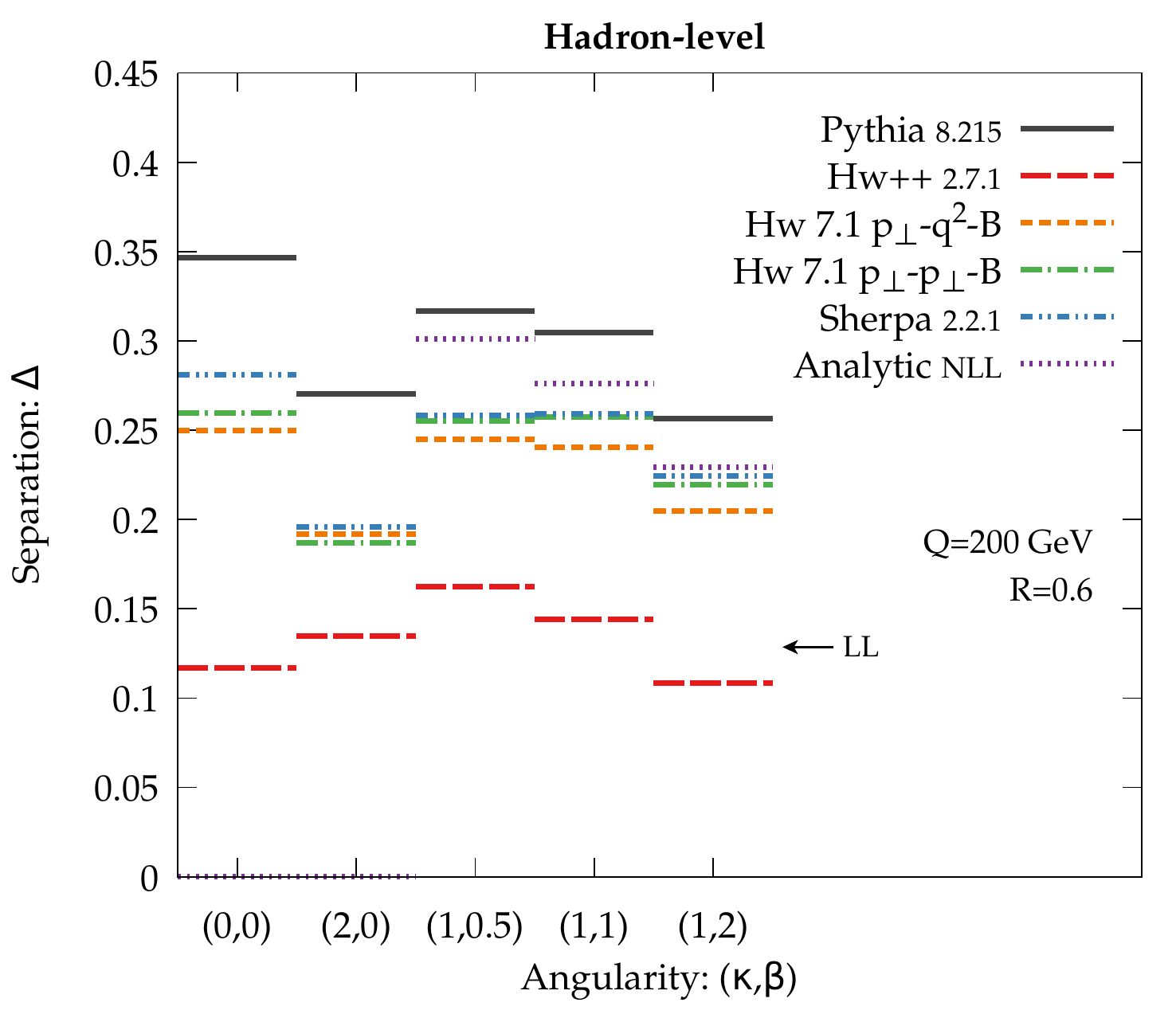}
}
\caption{Classifier separation $\Delta$ for the five angularities, determined from the various generators at hadron level
for an idealized case of $e^+ e^-$ collisions.  
The first two columns correspond to IRC-unsafe distributions (multiplicity and $p_T^D$), while the last three columns are the IRC-safe angularities.  
}
\label{fig:ee}
\end{figure}

\begin{figure}
\centering
{
\includegraphics[width = 1.0\columnwidth]{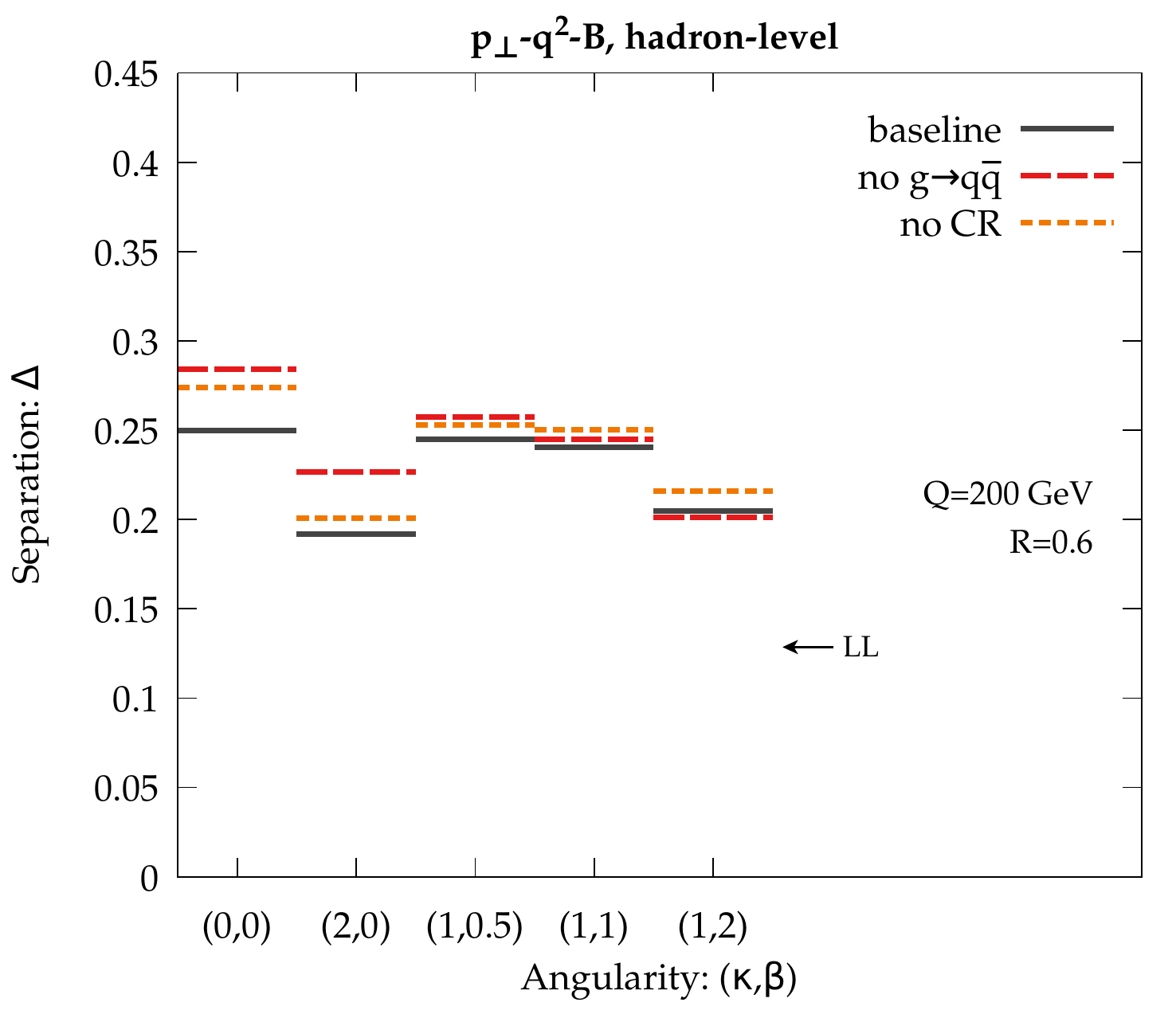}
}
\caption{Settings variations for Herwig 7.1 $p_\perp$-$q^2$-B tune.  Hadron-level results for the classifier separation $\Delta$ derived from the five benchmark angularities.}
\label{fig:ee_variation_herwig}
\end{figure}

We start with an idealized case of $e^+ e^-$ collisions (see Section 5 of~\cite{Gras:2017jty} for details). 
In Fig.~\ref{fig:ee} we show the discrimination power as a function of an angularity predicted by 
\textsc{Pythia 8.215} \cite{Sjostrand:2014zea},
\textsc{Herwig++ 2.7.1} \cite{Bahr:2008pv}, \textsc{Sherpa 2.2.1} \cite{Gleisberg:2008ta}
, the NNL analytical calculation from~\cite{Gras:2017jty} and the both $p_\perp$-$q^2$-B and 
$p_\perp$-$p_\perp$-B tunes of \textsf{Herwig~7.1}. Firstly, we see that the both \textsf{Herwig~7.1} 
tunes give significantly different results 
compared to \textsc{Herwig++ 2.7.1}. In order to understand the source of the difference, in Fig~\ref{fig:ee_variation_herwig} 
we investigate, for $p_\perp$-$q^2$-B tune, the following settings variations:
\begin{itemize}
\item \textsc{Herwig: no $g \to q\bar{q}$}.  Turning off $g \to q \bar{q}$ splittings in the parton shower.
\item \textsc{Herwig: no CR}.  The variation turns off color reconnections. 
\end{itemize}
We can see that the results are not very sensitive to the change of the settings.
This was not the case for \textsc{Herwig++ 2.7.1}
where the colour reconnection had a huge effect on the discrimination power, see ~\cite{Gras:2017jty}. 
Therefore, we can conclude that the difference is due to the improvements of the CR model 
described in Section~\ref{sec:hadron}, which as expected reduce effects of CR in the case of $e^+ e^-$ collisions.
Secondly, the results of the both \textsf{Herwig~7.1} tunes are quite similar and closer to the other predictions
giving more constrained prediction on the quark/gluon jet discrimination power in $e^+ e^-$ collisions. 
In fact just before finishing this paper the new tune was used
in~\cite{Mo:2017gzp}
confirming that indeed that improvements introduced in the manuscript
reduced
the tension between Pythia and Herwig and bring Herwig results closer to
NNLL'
results from~\cite{Mo:2017gzp}.

\begin{figure}
\centering
{
\includegraphics[width = 1.\columnwidth]{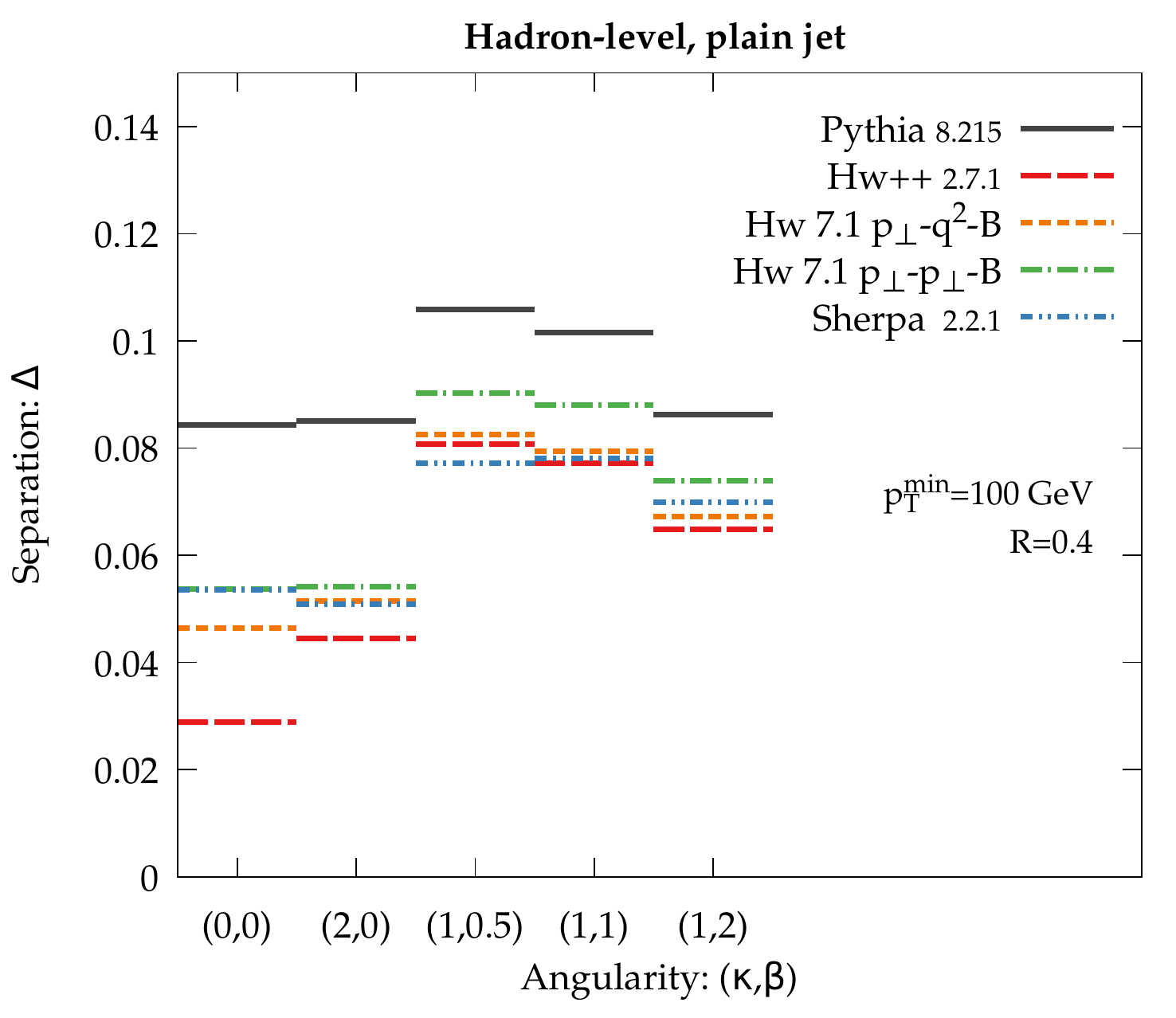}
\caption{Classifier separation $\Delta$ for the five angularities, determined from the various generators at hadron level
in the case of quark/gluon tagging at the LHC (see Section 6 of~\cite{Gras:2017jty} for details).  
The first two columns correspond to IRC-unsafe distributions (multiplicity and $p_T^D$), while the last three columns are the IRC-safe angularities.  
}
\label{fig:summary_hadron_pp_all}
}
\end{figure}

Next, in Fig.~\ref{fig:summary_hadron_pp_all} we show the results for $\Delta$ in the case of 
quark/gluon tagging at the LHC (see Section 6 of~\cite{Gras:2017jty} for details). Here we can see that
the differences between~\textsc{Herwig++ 2.7.1} and the both \textsf{Herwig~7.1} tunes are more modest when compared to the previous case of $e^+ e^-$ collisions.
However, as expected the largest differences between generators appear for IRC-unsafe observables
like multiplicity (0,0) and $p_T^D$ (2,0), where nonperturbative hadronization
plays an important role. It is also worth to notice that the $p_\perp$-$p_\perp$-B tune
which is preferred by the data on the charged particle multiplicity
in gluon jets at LEP~Fig.\,\ref{fig:gluonN}, and in jets at the LHC~Figs.\,\ref{fig:ATLASNCharged},\ref{fig:ATLASNdiff} 
gives slightly better discrimination power reducing the gap between predictions of \textsf{Pythia} and the other generators. 
Finally, it would be interesting to estimate the parton-shower 
uncertainties~\cite{Bellm:2016rhh,Bellm:2016voq,Mrenna:2016sih,Bothmann:2016nao}
in the context of the quark and gluon jet discrimination observables to see whether the remaining 
discrepancy in the predictions is covered by the uncertainty band.



 
\bibliographystyle{spphys}       
\bibliography{herwig}   

\end{document}